\documentclass[hyper]{JHEP3}

\usepackage{epsfig}
\usepackage{amsmath}
\usepackage{cancel}
\usepackage[bottom]{footmisc}

\newcommand{\definmath}[2] {\newcommand{#1}{\ensuremath{#2}}}

\newcommand{\slashchar}[1] {\cancel{#1}}

\definmath{\etmiss}{\slashchar{E}_T}
\definmath{\ptmiss}{\slashchar{p}_T}

\definmath{\Mbh}{M_{\text{BH}}}
\definmath{\Mpl}{M_{\rm PL}}
\newcommand{\compcode}[1]{\texttt{#1}}

\newcommand{\CHARYBDIS}{\compcode{CHARYBDIS}}

\definmath{\sumet}{\sum\!p_T}

\newcommand{\subfig}[3][]{\begin{minipage}{7cm}#3\vspace{-1em}\begin{center}{\small(#2)#1}\end{center}\end{minipage}}

\title{
Exploring Higher Dimensional Black Holes at the Large Hadron Collider
}

\author{C.M.~Harris$^\dag$, M.J.~Palmer$^\dag$, M.A.~Parker$^\dag$, P. Richardson$^{\ddag}$, A.~Sabetfakhri$^\dag$ and B.R.~Webber$^{\dag}$
\\
\\
$^\dag$ Cavendish Laboratory, University of Cambridge, Madingley Road,
Cambridge, CB3~0HE, UK.
\\
$^\ddag$ Institute for Particle Physics Phenomenology, University of Durham, DH1 3LE, UK.
}

\abstract{
In some extra dimension theories with a TeV fundamental Planck scale, black holes could be produced in future collider experiments.  Although cross sections can be large, measuring the model parameters is difficult due to the many theoretical uncertainties.  Here we discuss those uncertainties and then we study the experimental characteristics of black hole production and decay at a typical detector using the ATLAS detector as a guide.  We present a new technique for measuring the temperature of black holes that applies to many models.  We apply this technique to a test case with four extra dimensions and, using an estimate of the parton-level production cross section error of 20\%, determine the Planck mass to 15\% and the number of extra dimensions to $\pm$0.75.
}


\preprint{Cavendish-HEP-04/29\\
ATL-COM-PHYS-2004-067
}

\begin{document}

\section{Introduction}

In recent theories with extra dimensions the fundamental Planck mass, \Mpl{}, can be as low as the TeV scale~\cite{Arkani-Hamed:1998rs,Antoniadis:1998ig,Randall:1999ee,Randall:1999vf} making the trans-Planckian regime accessible for future high energy colliders. These models aroused great theoretical interest because they address the weak-Planck scale hierarchy. In these theories microscopic quantum black holes could be produced at energies higher than the Planck mass at the LHC~\cite{Argyres:1998qn,Banks:1999gd,Emparan:1999wa,Giddings:2001bu,Dimopoulos:2001hw,Hossenfelder:2001dn,Voloshin:2001vs, Lester:Talk}. Once produced, the black hole would decay very rapidly to a spectrum of particles by Hawking radiation. Assuming that all the Standard Model matter and gauge fields are confined to the physical three-branes in a higher dimensional space, it has been shown that most of the black hole decay products are Standard Model quanta emitted on the brane~\cite{Emparan:1999wa} and are therefore visible experimentally as very spectacular events. 

We stress that quantum extra dimensional black holes in no way constitute any threat, being distinguished from the more familiar astrophysical variety by being much lighter and highly unstable.  The astrophysical variety is much too heavy to be produced in current or planned collider experiments.  Hereafter, all discussion of black holes relates only to the extra dimensional variety.

In the large extra dimensions scenario~\cite{Arkani-Hamed:1998rs,Antoniadis:1998ig} at distances small compared with the size of the extra dimensions and in the warped scenario~\cite{Randall:1999ee,Randall:1999vf} at distances small compared to the curvature scale of the geometry associated with the extra dimensions, black holes with a horizon radius, $r_{\rm BH}$, smaller than the size of the extra dimensions can be treated as higher-dimensional objects located on the brane and extending along the extra dimensions. It has been shown that these small black holes have modified properties, e.g. they are larger and colder compared to a 4-dimensional black hole with exactly the same mass~\cite{Argyres:1998qn}.

The black hole discovery potential is critically dependent on the value of \Mpl{}. Short scale gravity experiments and particle collider experiments provide limits on the fundamental Planck scale. However for smaller values of the number of extra dimensions, $n$, the more stringent constraints come from astrophysical and cosmological data, albeit with larger uncertainties. It is widely agreed that the one large extra dimension scenario is ruled out by such data. The present collider limits\footnote{Limits are given in the convention of~\cite{Dimopoulos:2001hw} which is used throughout this paper (see section~\ref{sec:bhdecay}).} on the Planck scale range from 1.3 TeV for $n = 2$ to 0.3 TeV for $n = 6$ arising from the production of real (from missing energy signatures) or virtual Kaluza-Klein gravitons at the Tevatron Run I and LEP II~\cite{Abreu:2000vk, Abbiendi:2000hh, Acosta:2002eq, Abazov:2003gp}.  For a comprehensive recent review of these constraints see, for example,~\cite{Hewett:2002hv}.  The LHC, with a centre of mass energy of 14 TeV, offers a good opportunity for black hole production if $\Mpl \sim$ TeV. The very large cross section for production of black holes not too much heavier than the fundamental Planck scale corresponds to a production rate of a few Hertz at the LHC design luminosity.

In the following sections, the process of the black hole production and decay is reviewed (section \ref{sec:bhdecay}), followed by a description of the \CHARYBDIS{}~\cite{Harris:2003db, Harris:2003eg} event generator (section \ref{sec:event}).  We then present a review of the principal theoretical uncertainties (section~\ref{sec:ModelUncertainties}) before moving on to experimental discussions. The characteristics of black hole decays are presented in section \ref{sec:character} followed by a discussion of the measurement of the black hole mass in section~\ref{sec:mass}. We then discuss ways of determining the Planck mass (section~\ref{sec:MPlanck}) and finally, in section \ref{sec:ExtraD}, we study methods of determining the number of extra dimensions.  Throughout, we have used the ATLAS fast simulation software~\cite{Std:Atlfast2.0} to give a description of a typical detector and we have used the full simulation~\cite{Std:FullSim} to verify the main results.

\section{Black hole production and decay}
\label{sec:bhdecay}
In the black hole event generator \CHARYBDIS{}, which has been used in these studies, the black hole production is treated as a semi-classical process (black hole mass, $\Mbh \gg \Mpl$) and it is assumed that the extra dimensions are large ($\gg r_{\rm BH}$). For black hole masses close to \Mpl{} this semi-classical approximation is not valid and a theory of quantum gravity would be required to calculate the cross section. To be within the semi-classical domain we restrict the mass of the black hole to be $\Mbh \ge 5 \, \Mpl$. By geometrical arguments the semi-classical parton-level cross section for black hole production would be~\cite{Banks:1999gd} 
\begin{equation}\label{eq:geometry}
   \hat{\sigma} (\hat{s} = \Mbh^2)\approx \pi \, r^2_{\text{BH}}
\end{equation}
where $\sqrt{\hat{s}}$ is the centre-of-mass energy of the colliding particles (see~\cite[equation 2.4]{Harris:2003db}).  

The radius for a Schwarzschild black hole is
\begin{equation}
r_{\rm BH} = \frac{1}{\sqrt{\pi}\Mpl}\left(\frac{\Mbh}{\Mpl}\right)^{\frac{1}{n+1}}\left(\frac{8\Gamma\left(\frac{n+3}{2}\right)}{n+2}\right)^{\frac{1}{n+1}}
\label{eq:radius}
\end{equation}  
where we have used the convention $\Mpl^{n+2} = 1/G_{(n+4)}$ where $G_{(n+4)}$ denotes the $n+4$ dimensional Newton's constant~\cite{Dimopoulos:2001hw}, so for a fixed black hole mass, the cross section is lower for a higher number of extra dimensions.  This convention has been used throughout this paper.

The decay of a spinning black hole comprises three phases~\cite{Giddings:2001bu}: 1) the \textit{balding phase}, in which the black hole loses its `hair' (associated with the multipole moments) by the emission of radiation; 2) the \textit{evaporation phase}, which starts with a brief spin-down phase, shedding away its angular momentum, followed by the \textit{Schwarzschild phase}, emitting a large number of quanta which reduce the mass of the black hole; 3) finally the \textit{Planck phase} (also called the \textit{remnant decay}), when \Mbh{} approaches \Mpl{}, in which the final decay takes place by the emission of a few quanta. 

\CHARYBDIS{} only models the Schwarzschild phase which is expected~\cite{Giddings:2001bu} to account for the greatest proportion of the mass loss.  The energy spectrum of decay products is approximately black body with corrective `grey-body' factors~\cite{Harris:2003eg}, $\gamma$, which the generator includes.  The spectrum for a \textit{fixed} temperature black hole is
\begin{equation}
\frac{dN}{dE} \propto \frac{E^2 \gamma}{\left(e^{E/T_H} \mp 1\right) T_H^{n+6}}
\label{eq:Spectrum}
\end{equation}
the denominator includes a spin-statistics term which is $-1$ for bosons and $+1$ for fermions.  The energy spectrum has a characteristic Hawking temperature, $T_H$, which is given by
\begin{equation}
T_H = \frac{n+1}{4\pi r_{\text{BH}}}
\label{eq:Temperature}
\end{equation}
and is thus related to the black hole mass and the number of extra dimensions by
\begin{equation}
\log T_H = \frac{-1}{n+1}\log \Mbh + \text{constant}
\label{eq:TempMass}
\end{equation}
where the constant is dependent only on $n$ and \Mpl{}.  The generator can also model the time dependence in which case $T_H$ is recalculated after each emission (so as the black hole decays it gets hotter).  Otherwise the initial $T_H$ is used throughout the decay.

Due to their high mass, black hole decays are very spectacular events with a large visible transverse energy, large multiplicity, and high sphericity with many hard jets and leptons. Since most of the black hole decay products result from the evaporation phase, as visible Standard Model particles, the ratio of the total hadronic to leptonic activity is expected to be roughly 5:1~\cite{Giddings:2001bu}. A few hard quanta with energy a sizable fraction of the \Mpl{} are also expected from the final Planck decay phase~\cite{Giddings:2001bu}. 

The theoretical work to date has been done in the semi-classical approximation.  This approximation is only valid if $\Mbh \gg \Mpl$, $\Mbh \gg T_H$ and the average multiplicity is large, $\langle N \rangle \gg 1$.  This approximation can only be valid at the LHC if the Planck mass is low and even then, there will be problems if the number of dimensions is large (since this gives a temperature close to the Planck mass and thus low multiplicity).

\section{Event generation and detector simulation}
\label{sec:event}

\CHARYBDIS{} has been used to generate Monte Carlo event samples. It is interfaced, via the Les Houches accord~\cite{Boos:2001cv}, to \compcode{HERWIG}~\cite{Corcella:2000bw,Corcella:2002jc} to perform the parton shower evolution of the partons produced in the decay and their hadronization.  The generated events are then passed through the ATLAS fast simulation, \compcode{ATLFAST}~\cite{Std:Atlfast2.0}, in order to give a reasonable description of detector resolution and efficiency.

Unless otherwise mentioned, the \CHARYBDIS{} options were set as follows: 
\begin{itemize}
\item{Time variation of the black hole temperature was on (\compcode{TIMVAR=TRUE}).}
\item{Grey-body effects were on (\compcode{GRYBDY=TRUE}).}
\item{The black hole was allowed to decay to all Standard Model particles including Higgs particles (\compcode{MSSDEC=3}).}
\item{Kinematic cut-off was turned off (\compcode{KINCUT=FALSE}).}
\item{The number of particles in the remnant decay was 2 (\compcode{NBODY=2}).}
\end{itemize}
This set of options together with the Planck mass set to 1~TeV is called the `test case' and we have used this to illustrate our techniques.  Several samples have been generated, so to avoid confusion the number of dimensions is always specified.  If a mass is given, then the generator was forced to produce black holes with a fixed mass, otherwise the range was set to 5000--14000~GeV.

Table~\ref{tab:crossbh} summarises the black hole production cross sections at the LHC for $n = 2, \,4$, and 6 with $\Mpl=1$~TeV.\footnote{In this analysis, we have used the MRSD-' (DIS) parton distribution function (PDF) set~\cite{Martin:1993zi} with $Q^2 = 1 / r^2_{\text{BH}}$, where $Q^2$ is the momentum scale squared at which a PDF is evaluated.} 

\TABLE{
\renewcommand{\arraystretch}{1.2}
\begin{tabular}{|l|c|c|}
  \hline
  \multicolumn{2}{|c|}{Topology} & Total Cross Section (fb) 
  \\ \hline \hline
                     & $n = 2$ &  $ 62,000 $ 
  \\ \cline{2-3}
   5 TeV black hole  & $n = 4$ &  $ 37,000 $ 
  \\ \cline{2-3}
                     & $n = 6$ &  $ 34,000 $  
  \\ \hline
                     & $n = 2$ &  $ 580 $   
  \\ \cline{2-3}
   8 TeV black hole  & $n = 4$ &  $ 310 $   
  \\ \cline{2-3}
                     & $n = 6$ &  $ 270 $   
  \\ \hline
                     & $n = 2$ &  $ 6.7 $    
  \\ \cline{2-3}
  10 TeV black hole  & $n = 4$ &  $ 3.4 $    
  \\ \cline{2-3}
                     & $n = 6$ &  $ 2.9 $    
  \\ \hline
\end{tabular}
\caption{\label{tab:crossbh} The black hole production cross sections at the LHC for $\Mpl=1$~TeV as given by \CHARYBDIS{}.  Note that \CHARYBDIS{} does not include the form factors mentioned in section~\ref{sec:MPlanck}.}
}

\section{Model uncertainties}
\label{sec:ModelUncertainties}

The theory of black hole production and decay contains many uncertainties and assumptions, particularly at LHC energies.  A clear understanding of these is therefore essential in order for our analyses to be as widely applicable as possible.  In this section we review these uncertainties.

\subsection{Production cross section}
\label{sec:CrossSection}

The process of black hole production in hadron collisions is
subject to a number of basic uncertainties.  The order of magnitude 
of the parton-level cross section should be given by equation~\ref{eq:geometry}, but the form factor relating the left- and right-hand sides
is uncertain and would be expected to be $n$-dependent. Classical
numerical simulations~\cite{Yoshino:2002tx} suggest values in the
range 0.5--2, increasing with $n$.  These values are not included
in the \CHARYBDIS{} generator, but we take them into account
when cross section data are used in our analysis (in sections \ref{sec:MPlanck} and \ref{sec:ExtraD}).

More fundamentally, the transition from the parton-level to the 
hadron-level cross section is based on the factorization formula
\begin{equation}
\sigma(S) = \int dx_1\,dx_2\,f(x_1)f(x_2)\hat\sigma(\hat s=x_1 x_2 S)
\end{equation}
where $f(x)$ is the parton distribution function (PDF) summed over
parton flavours.  The validity of this formula in the trans-Planckian
energy region is unclear.  Even if factorization remains valid,
the extrapolation of the PDFs into this region based on Standard Model
evolution from present energies is questionable.  Also, comparison to Standard Model processes in the trans-Planckian regime would be difficult since perturbative physics would be suppressed.

\subsection{The first stages of decay}
\label{sec:EarlyStages}
\CHARYBDIS{} does not model the initial balding or spin-down phases of the black hole decay.  The amount of energy emitted from the black hole during these phases is expected to be small~\cite{Giddings:2001bu} so such an omission should not be significant.  However, it is probable that the energy spectrum will be modified at low energies.

\subsection{Deposition on the brane}
Estimates vary as to how much energy is expected to be emitted into the bulk via graviton emission, but it could be significant.  One estimate suggests that the fraction of energy emitted into the bulk could be as high as 20\% for $n=2\text{--}4$ rising to nearly 50\% for $n=7$~\cite{Harris:Thesis}.  Any energy emitted into the bulk will make an accurate measurement of the mass extremely difficult.  Although this effect could in principle be observed as a change in the expected shape of the cross section as a function of black hole mass, determining this would be experimentally challenging.  In these studies we have assumed that all the energy is deposited on our brane.  A modified generator and a more detailed study would be necessary to understand the full impact of this assumption.

\subsection{Kinematic limit}
\label{sec:KinematicLimit}

A black hole can only emit a particle with an energy up to half of its mass in order to conserve energy--momentum. However, the grey-body distribution used to describe the Hawking radiation extends to infinite energy.  Although the distribution is only valid for very massive ($\Mbh\gg \Mpl$) black holes it is still necessary to deal with the black holes as they become lighter. It is expected that given a full theory, the distribution would be modified to take this into account, but we have no such theory.

Figure~\ref{fig:GenCorr} shows the energy of the primary generator level decay products in the rest frame of the black hole. As can be seen, the kinematic limit affects most of the decays. This greatly modifies the energy spectrum and also leaves open the question of what to do when the generator chooses an unphysical decay, i.e. when it samples from the energy spectrum above the kinematic limit. Two options are implemented in \CHARYBDIS{}.  In the first case (\compcode{KINECUT=FALSE}), if an unphysical decay is chosen, it is thrown away and a new one is chosen.  This process continues until the black hole has a mass less than the Planck mass at which point the decay moves to the final, remnant, stage.  In the other option (\compcode{KINECUT=TRUE}), when an unphysical decay is chosen, the black hole decay moves straight to the final stage.  The final stage of the decay is dealt with in the next section.  It should be noted that for high temperature black holes, where the probability of an unphysical decay is large, the different choices implemented in the generator will lead to a large difference in the multiplicities and will have a significant impact of the energy distributions.

\FIGURE{
\subfig{a}{\epsfig{file=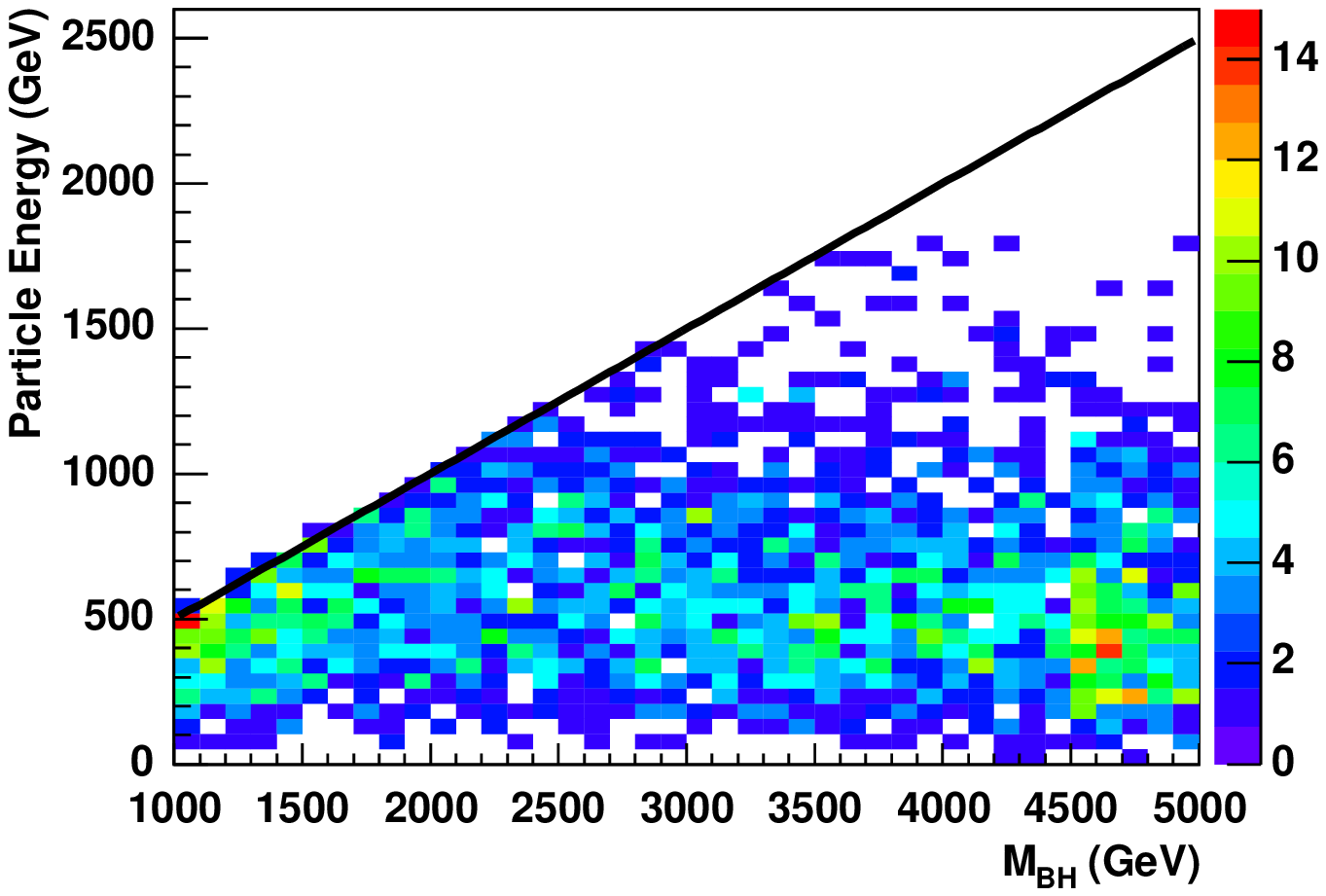, width=7cm}}
\subfig{b}{\epsfig{file=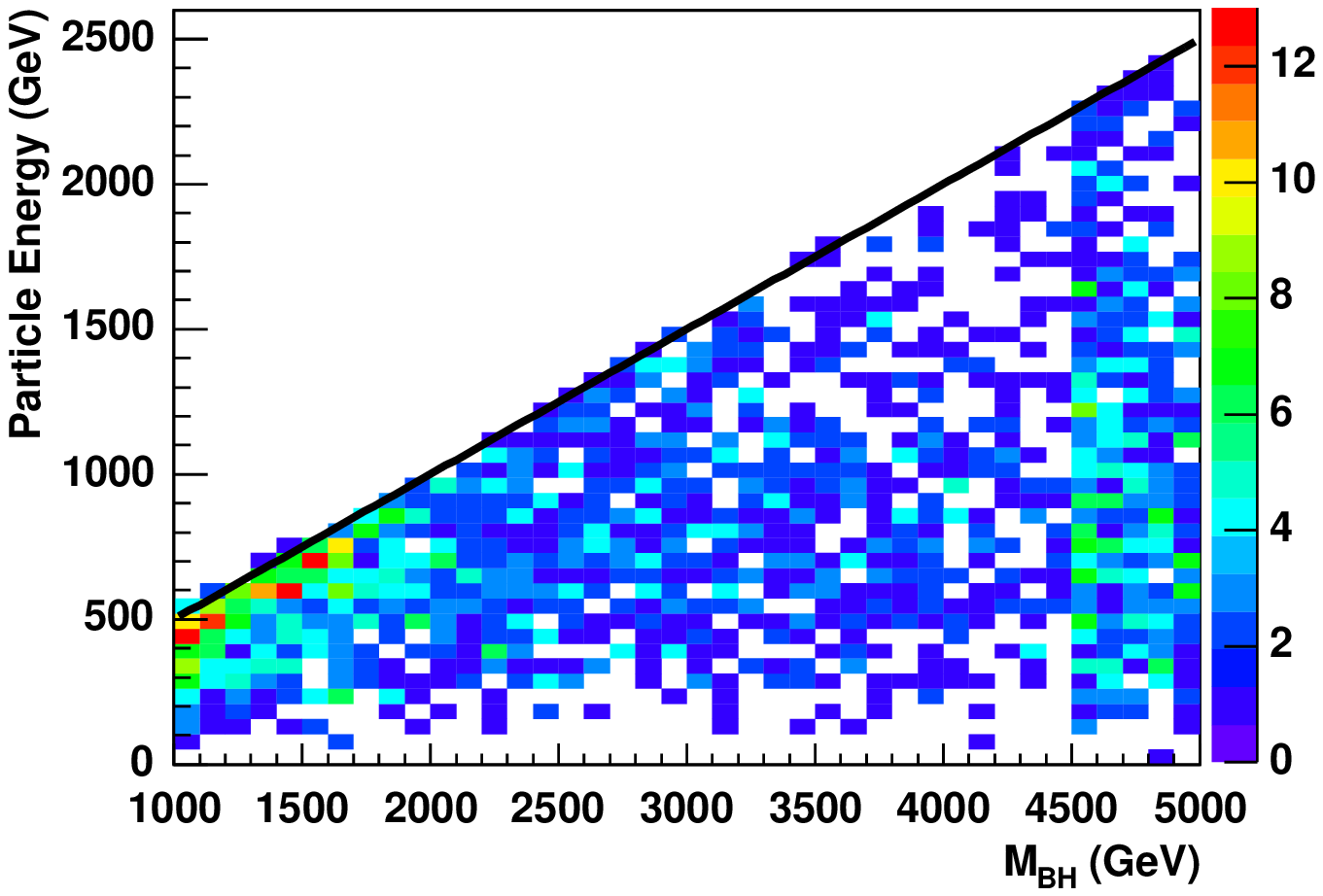, width=7cm}}
\caption{
Energy of the generator level decay products in the rest frame of the black hole for a 5~TeV black hole and 1000 events. The colour scale indicates the number of particles in each bin. (a) for $n=2$ the kinematic limit ($E=\Mbh/2$, black lines) constricts the energy distribution at low masses. (b) for $n=4$ the kinematic limit clearly affects the energy distribution at all masses.
}
\label{fig:GenCorr}
}

\subsection{Remnant decay}
\label{sec:RemnantDecay}

At the end of the evaporation phase, a light Planck scale black hole, called a remnant, remains which the generator must decay. How this would happen is unknown and will only be predicted by a quantum theory of gravity. \CHARYBDIS{} implements this `remnant' decay as an isotropic decay into 2--5 bodies (the number is an option). When the remnant decay occurs depends on the option chosen for handling the kinematic limit as described in section~\ref{sec:KinematicLimit}.  It should be noted therefore, that the uncertainty here can easily affect the multiplicity and energy spectra.  One example of an affected experimental observable is the photon energy spectrum.  Figure~\ref{fig:nbody} shows the photon energy distributions (of all photons in the event) for 2-body and 4-body remnant decays for two values of $n$.  Even for $n=2$ there is a noticeable effect, but for $n=4$, the effect is large. 

\FIGURE{
\subfig{a}{\epsfig{file=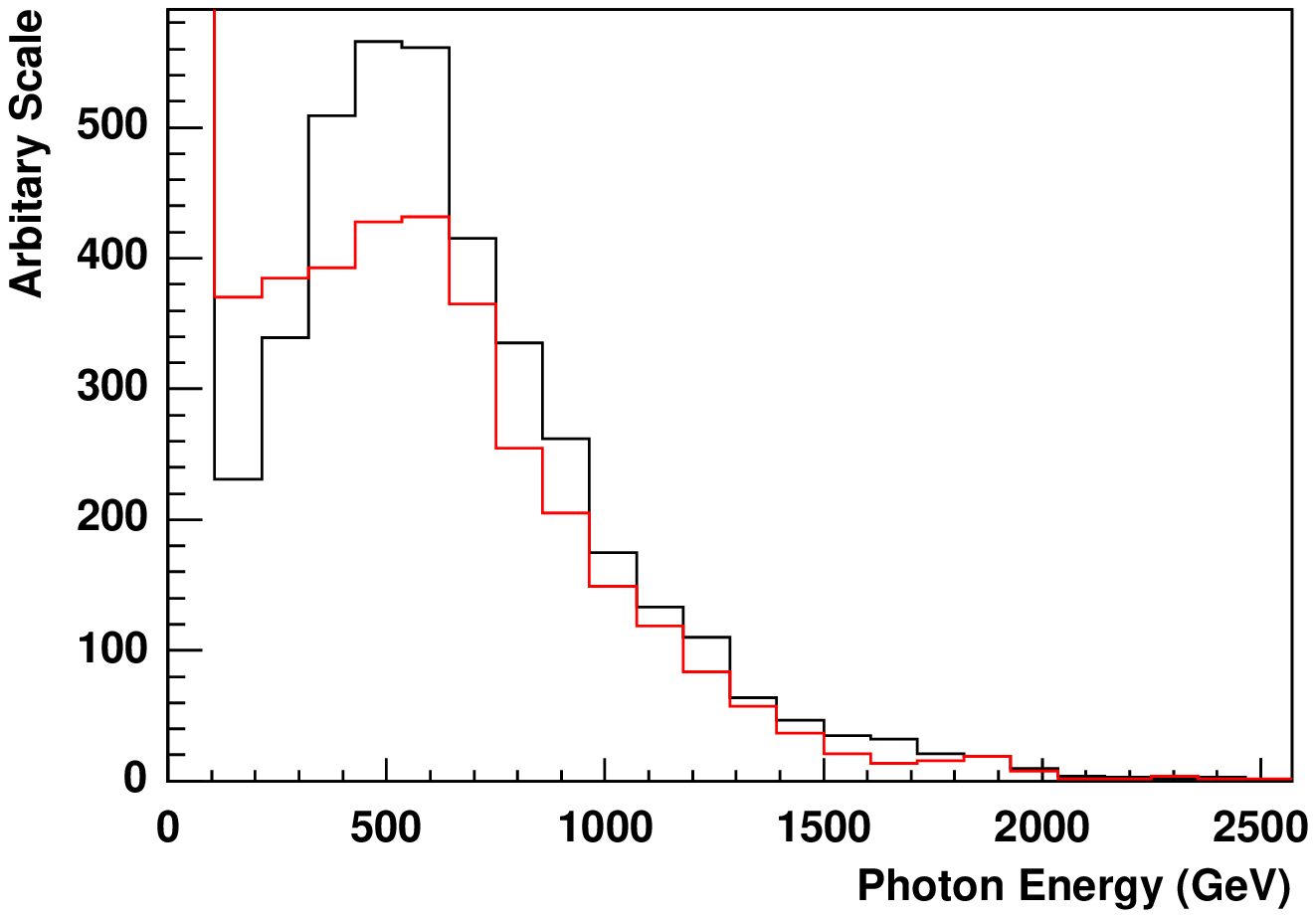, width=7cm}}
\subfig{b}{\epsfig{file=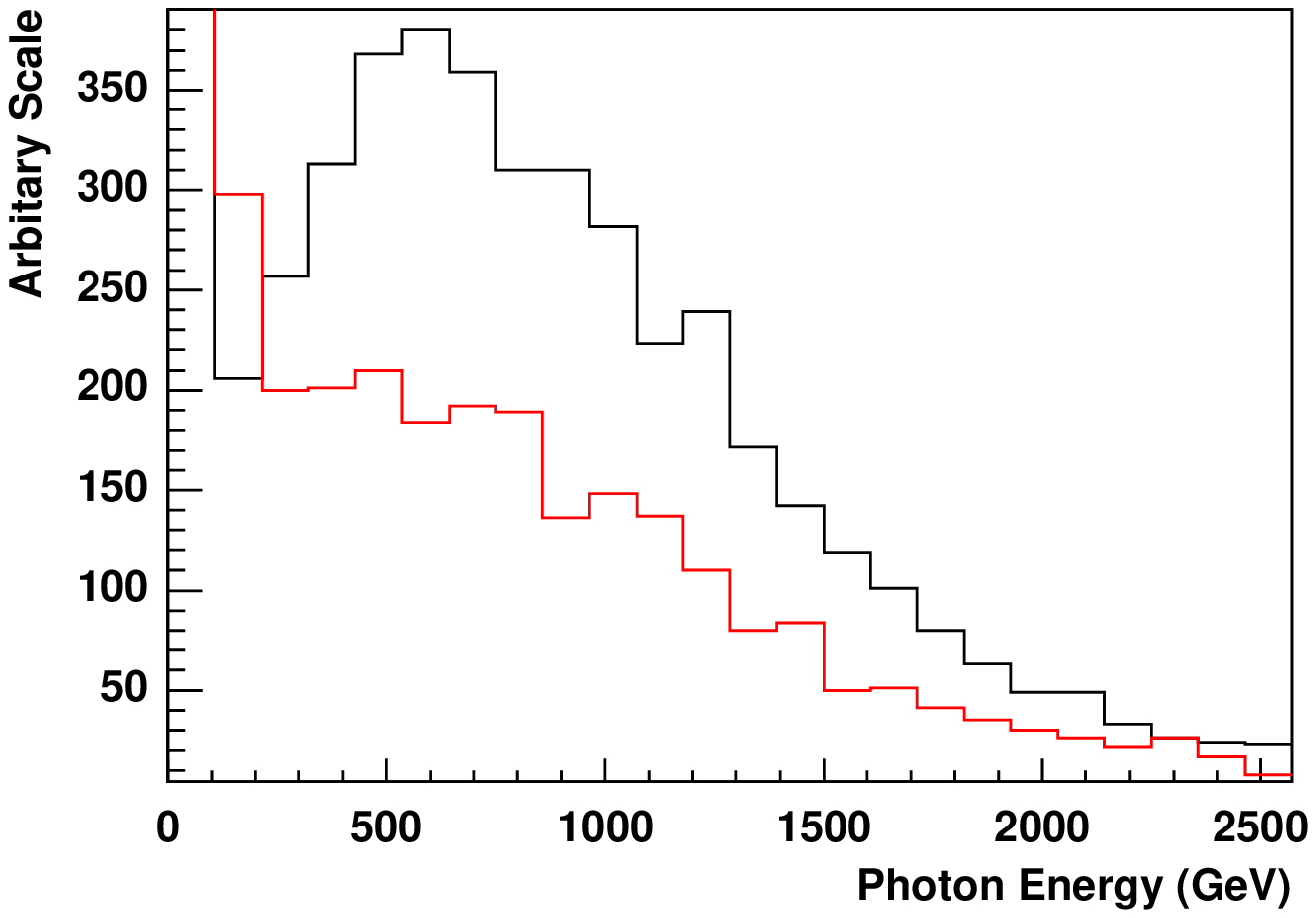, width=7cm}}
\caption{The photon energy distributions for (a) $n=2$ and (b) $n=4$.  The black and red lines are for 2-body and 4-body remnant decays respectively.}
\label{fig:nbody}
}

\subsection{Time-variation and black hole recoil}
\label{sec:TimeVar}

It has been argued~\cite{Dimopoulos:2001hw} that due to the speed of the decay, the black hole does not have enough time to equilibrate between emissions and therefore that the time variation of the temperature can be ignored. Therefore, the initial Hawking temperature might be measured by fitting Planck's formula for black-body radiation to the energy spectrum of the decay products for different bins in the initial black hole mass. Using equation~\ref{eq:TempMass} the number of dimensions can then be extracted.  This is the approach taken at a theoretical level in~\cite{Dimopoulos:2001hw}.

To illustrate this procedure, we have used the test case with $n = 2$.  Events were generated without grey-body factors in 500~GeV mass bins between 5000 and 10000~GeV.  For each mass bin we have fitted the black-body spectrum to the generator level electron energy.  Figure~\ref{fig:tempnot}a shows the result of this together with the fit using equation~\ref{eq:TempMass} from which we determine $n=1.7\pm 0.3$. Figure~\ref{fig:tempnot}b shows the result of the same procedure and the same test case but with time dependence turned on.  In this case we determine $n=3.8 \pm 1.0$ which is well away from the model value.  Time dependence is therefore a systematic effect with a strong impact on any measurement of $n$.

\FIGURE{
\epsfig{file=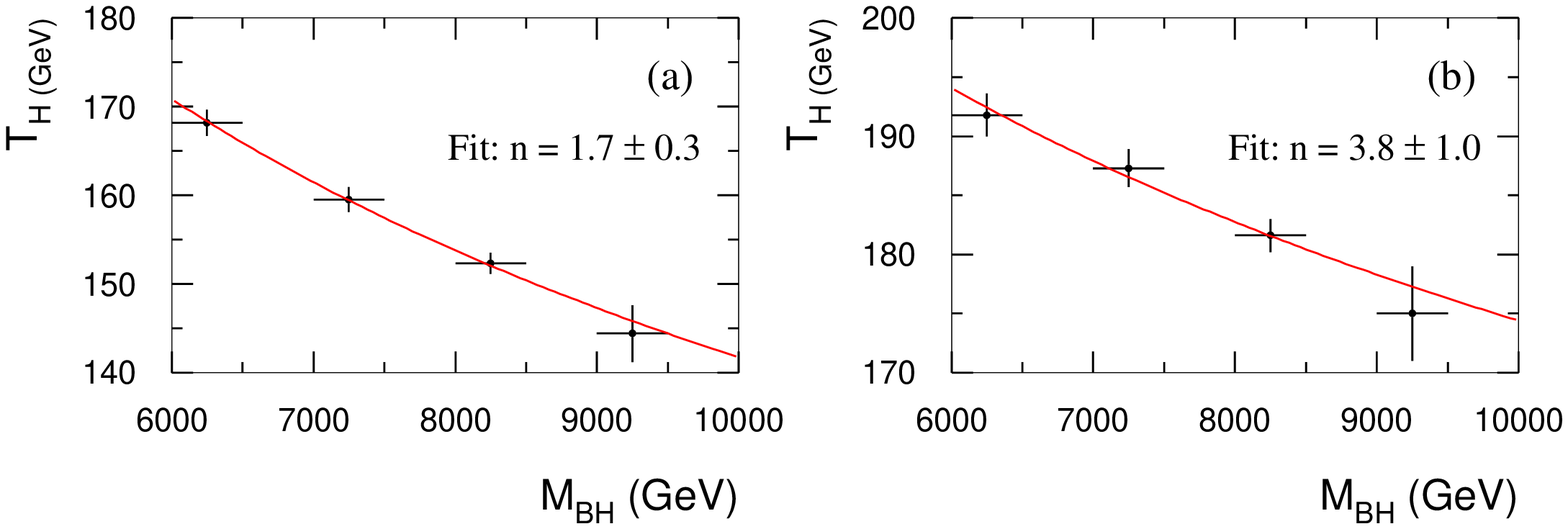,width=\textwidth}
\caption{\label{fig:tempnot}The plot of $T_H$ versus \Mbh{} for $n = 2$ and $\Mpl = 1$~TeV, (a) with a fixed Hawking temperature, and (b) with changing (time dependent) Hawking temperature. The statistics used correspond to 30~fb$^{-1}$ of running at the LHC.}
}

Another effect that has not been taken into account in previous studies is the recoil of the black hole.  When a particle is emitted from the black hole, the black hole recoils against it.  Therefore the next emission is in a boosted frame.  Even in the case of a fixed temperature decay, the effects of recoil become more significant as the decay progresses and the black hole gets lighter.  This is exacerbated in the time varying case since the black hole also gets hotter as it decays.  Any analysis which makes use of the energy spectrum should therefore account for this.

\section{Characteristics of the black hole decay}
\label{sec:character}

Black hole decays in the semi-classical limit have high multiplicity.  However at LHC energies black holes would be on the edge of the semi-classical limit (depending on $n$) which can reduce the multiplicity and make predictions uncertain.  This effect can be seen in figure~\ref{fig:multiplicity} which shows that the multiplicity decreases significantly with $n$.  This is due to fact that $T_H$ is higher for larger $n$ at the same mass. 

\FIGURE{
\subfig{a}{\epsfig{file=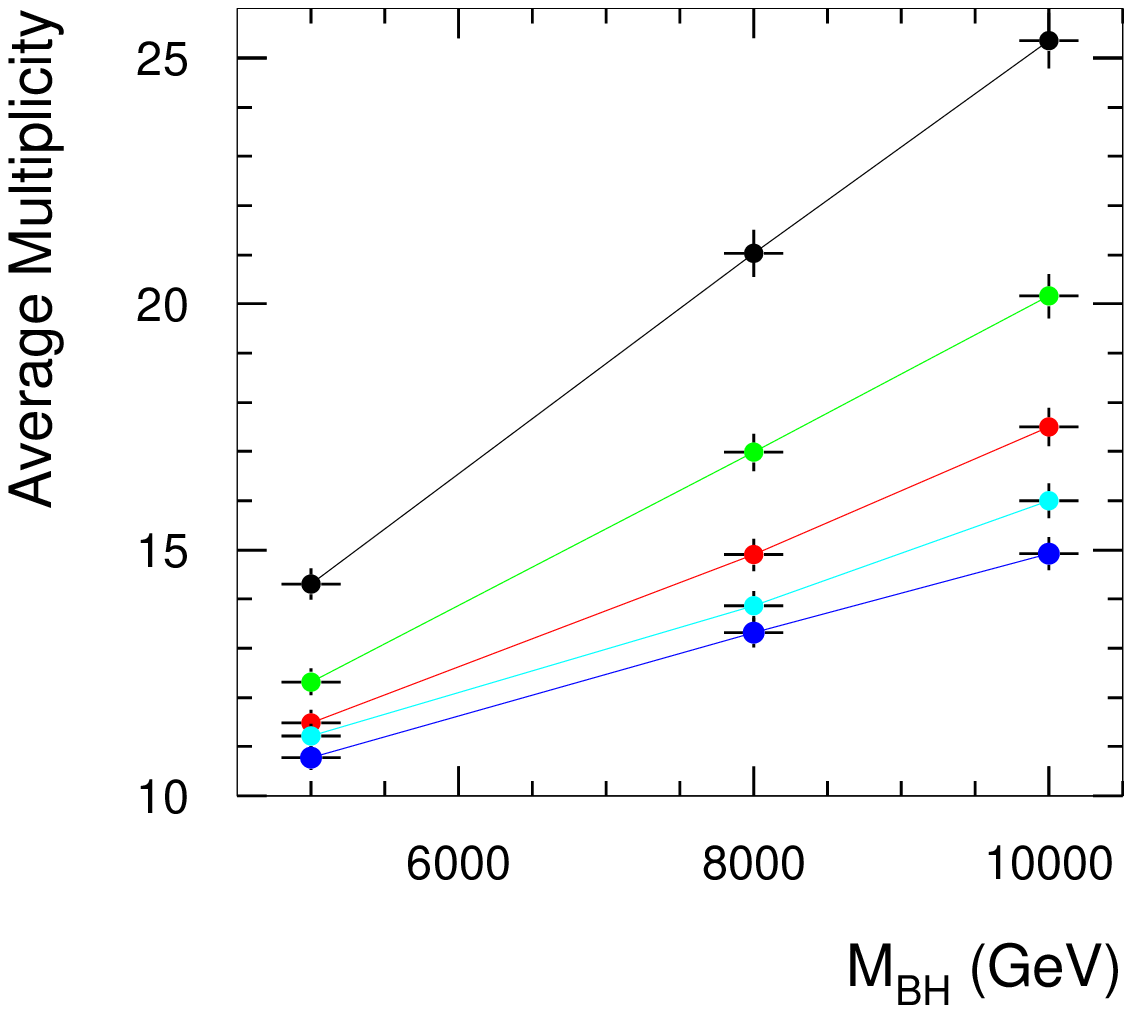, width=7cm}}
\subfig{b}{\epsfig{file=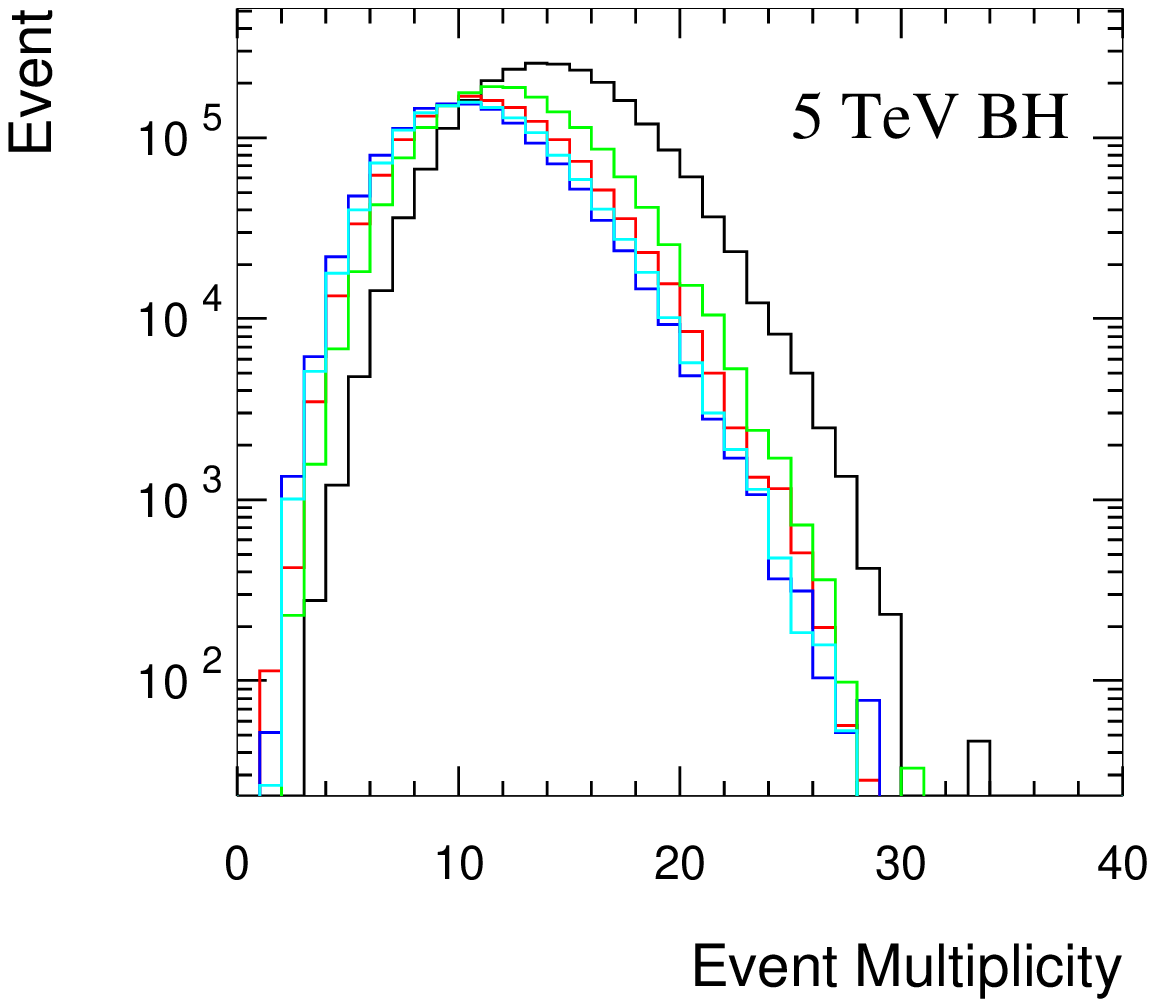, width=7cm}}
\caption{(a) Average event multiplicities and (b) typical multiplicity distribution with $n=2, \, 3, \, 4, \, 5,$ and 6 (black, green, red, cyan, and blue curves, respectively) for 100~fb$^{-1}$ of integrated luminosity.}
\label{fig:multiplicity}
}

A black hole decay is also characterised by a large total transverse energy (figure~\ref{fig:sumet}) which increases as the black hole mass increases.  Even the low multiplicity events tend to be rather spherical with high multiplicity events more so.  These characteristics are very different from standard model and SUSY events which do not have the same access to very high energies and tend to produce less spherical events.  Therefore, we believe that selecting events with high \sumet{}, high multiplicity ($>4$) and high sphericity will give a pure set of black hole events.  In addition, it should be noted that the already small Standard Model background will be suppressed by the black hole production~\cite{Dimopoulos:2001hw}.  There are two further characteristics which will be interesting to measure and confirm the nature of the events: the missing $p_T$ (\ptmiss{}) distribution and the charge asymmetry.

\FIGURE{
\epsfig{file=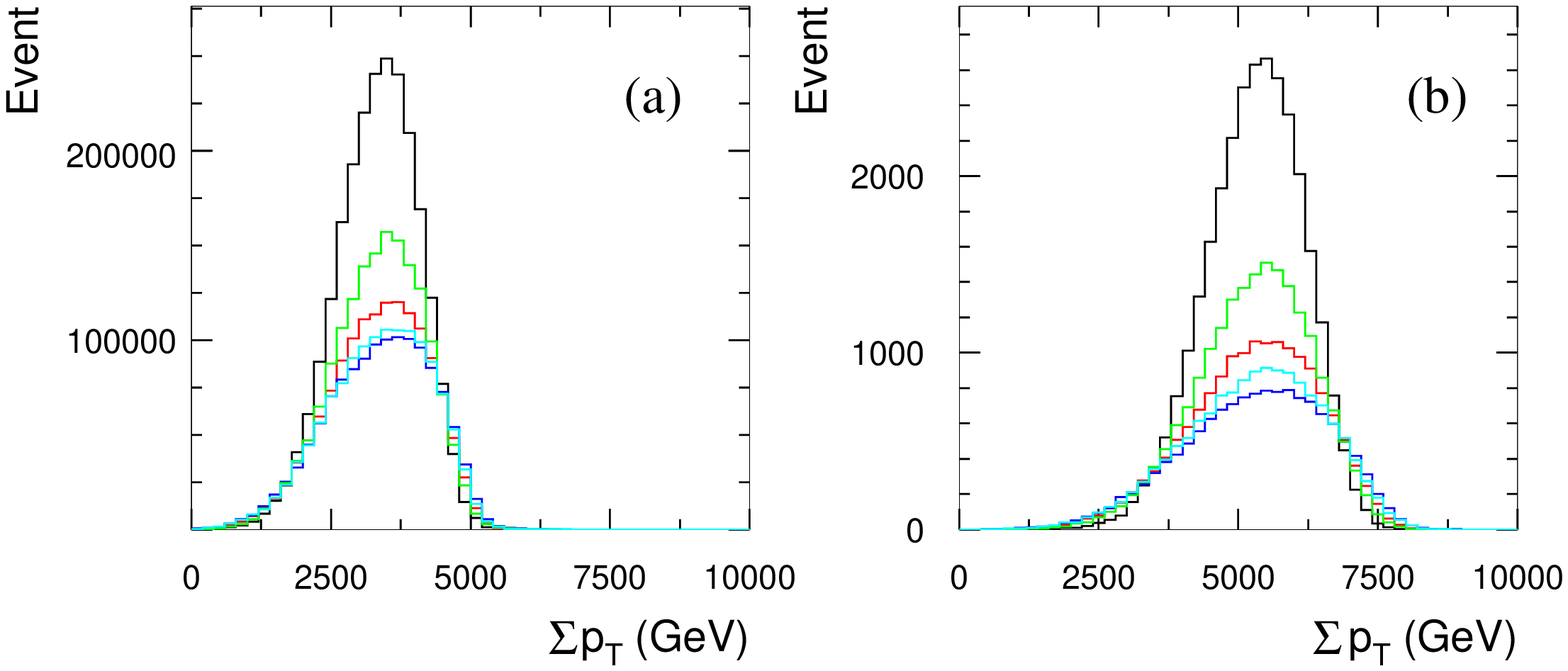,width=\textwidth}%
\caption{The distribution of \sumet{} for $n=2, \, 3, \, 4, \, 5,$ and 6 (black, green, red, cyan, and blue curves, respectively) for 100~fb$^{-1}$ of integrated luminosity.  (a) for 5~TeV and (b) for 8~TeV black holes.}
\label{fig:sumet}
}

\subsection{\ptmiss{} distribution}

Although not all black hole decays contain neutrinos, some will have one or more with energies that can be as high as half the black hole mass.  The missing energy can be even larger than for much of SUSY parameter space. In contrast, most of the Standard Model processes tend to have much lower missing transverse momenta. Figure~\ref{fig:missptcom} presents the distribution of the \ptmiss{} for Standard Model QCD events (with generator level cut $p_T >$ 600~GeV), SUSY events (at LHCC SUGRA point 5~\cite{LHCC:SUGRA}), and 5~TeV black holes with $n=2$ and 6.

\subsection{Black hole charge}

Black holes are typically formed from valence quarks, so it is expected that the black holes would be charged.  The average charge is somewhat energy dependent, but should be $\sim +2/3$.  The rest of the charge from the protons is expected to disappear down the beam pipes or at very high $|\eta|$.  The average black hole charge, $\langle Q_{\text{BH}}\rangle$, can be measured by determining the average charge of the charged leptons, $\langle Q_{\text{Lept}}\rangle$, which should be equal to the black hole charge times the probability of emitting a charged lepton.  Figure~\ref{fig:leptonCharge} shows such a measurement for the test case with $n=2$ which gives $\langle Q_{\text{Lept}}\rangle = 0.1266\pm0.002$ and thus $\langle Q_{\text{BH}}\rangle = 0.654\pm0.008$ using the expected charged lepton emission probability of 0.1936.

\DOUBLEFIGURE{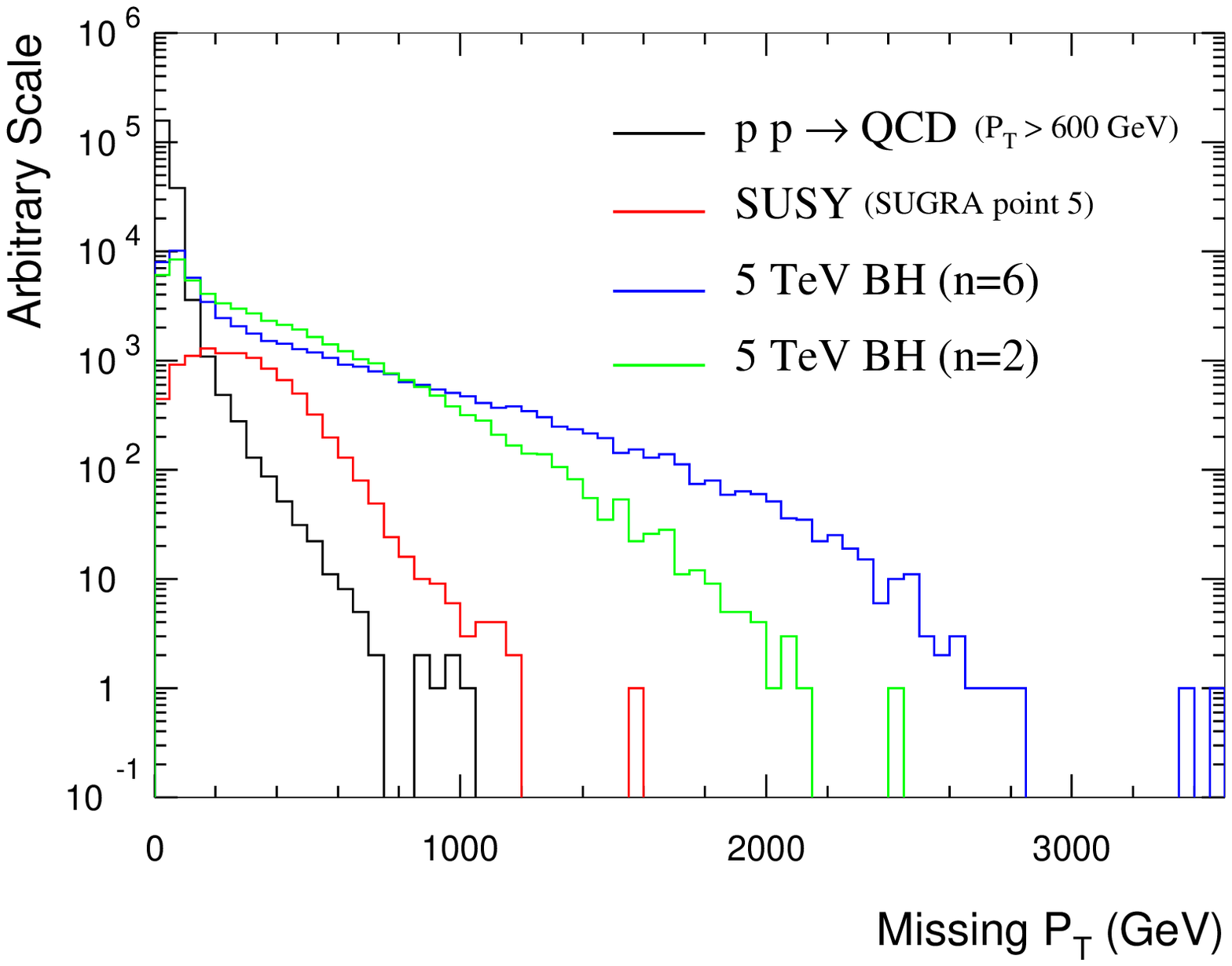, width=0.53\textwidth}{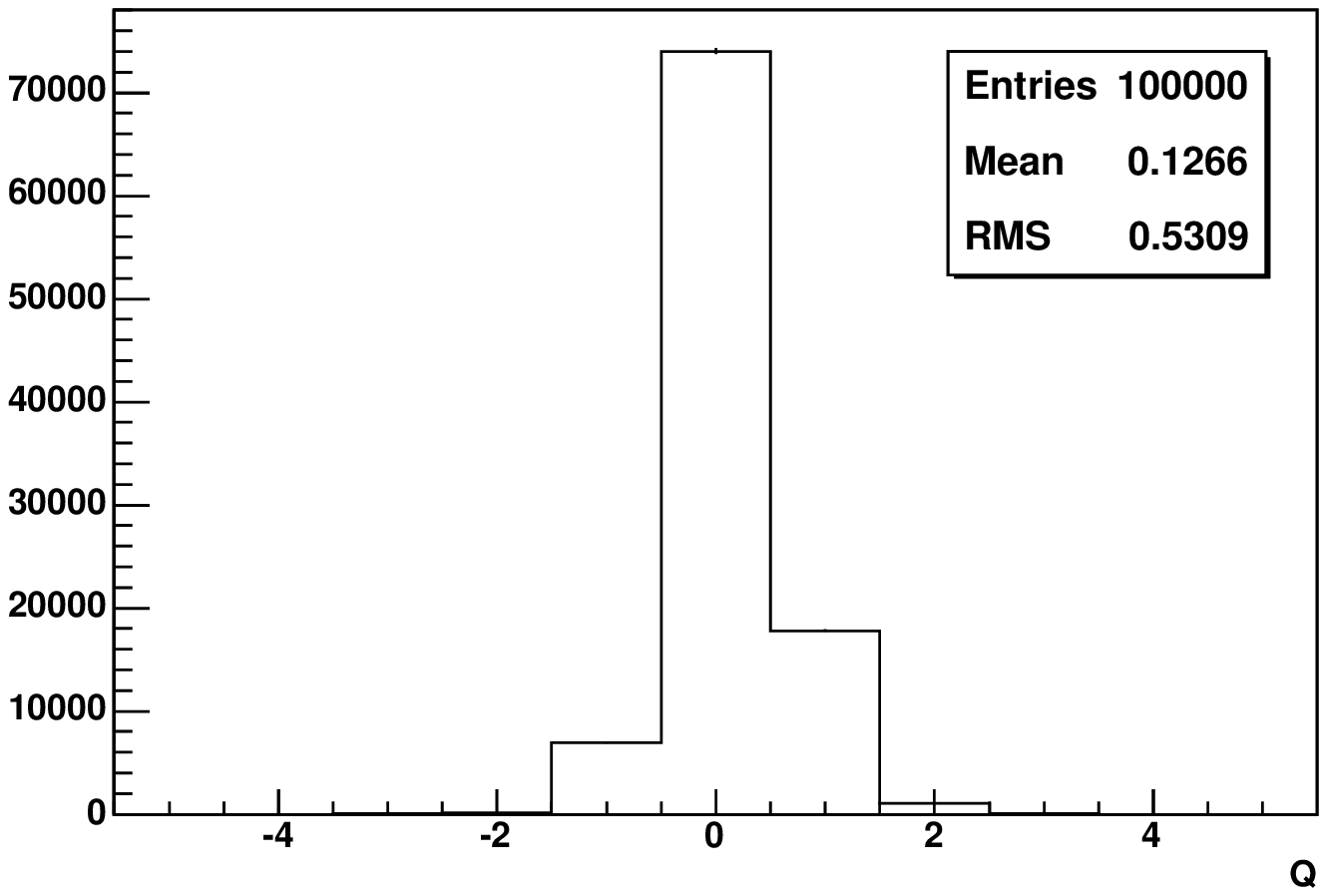, width=0.47\textwidth}{\label{fig:missptcom} The distribution of the \ptmiss{} for Standard Model QCD events (with generator level cut $p_T >$ 600~GeV), SUSY events (at LHCC SUGRA point 5), and 5~TeV black hole with $n=2$ and 6.}{\label{fig:leptonCharge}The average charge of electrons and muons for $n=2$ with approximately $1\text{~fb}^{-1}$ of data.}

\subsection{Kinematic distributions}
\label{sec:spectra}

The authors of~\cite{Mocioiu:2003gi} have studied the hadronic decay of a black hole and found that the transverse momentum distribution of charged hadrons depends weakly on the number of large extra dimensions.  In addition to the event multiplicity and transverse momentum distribution, figure~\ref{fig:pt8}, we have also looked at the average $p_T$ of the events, jets, leptons, and the ratio of the difference and sum of the $i^{th}$ and the $j^{th}$ highest $p_T$ jet ($i \, ,j = 1, \, 2, \, 3, \, 4$) and found that these variables also depend only weakly on $n$.  It is therefore not possible to get a constraint on $n$ using these distributions.

\FIGURE{
\epsfig{file=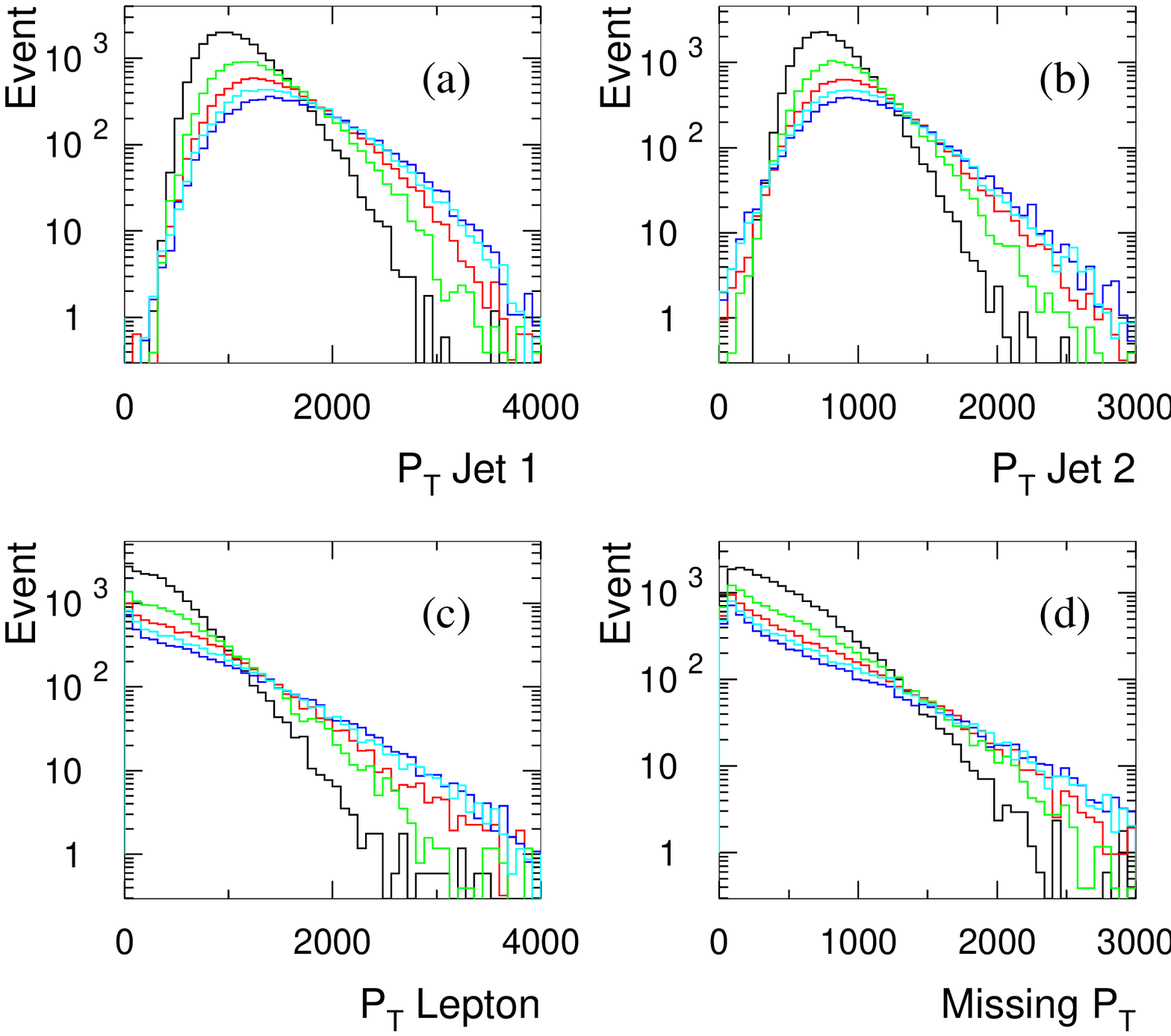,width=0.9\textwidth}
\caption{\label{fig:pt8} The distribution of (a) the transverse momenta for the highest $p_T$ jets, (b) the second highest $p_T$ jets, (c) the $p_T$ of the leptons, and (d) \ptmiss{} in the decay of 8~TeV black holes with $n=2, \, 3, \, 4, \, 5,$ and 6 (black, green, red, cyan, and blue curves, respectively) for 100~fb$^{-1}$ of integrated luminosity.}
}

\subsection{Event shape variables}
\label{sec:shape}

In addition to the event multiplicity and spectra, we have studied the following event shape variables: the sphericity~\cite{Bjorken:1970wi}, thrust~\cite{Brandt:1964sa}, and the Fox-Wolfram moment ratios~\cite{Fox:1979id}. Since the sphericity ($S$) and thrust ($T$) are sensitive to underlying event and longitudinal motion, we have used the corresponding quantities for transverse momenta only.

Defining the Fox-Wolfram moments~\cite{Fox:1979id}
\begin{equation}\label{eq:fox}
     H_l = \sum_{i,j} \frac{\left| \mathbf{p}_i \right|
   \left| \mathbf{p}_j \right|}{E^2_{\rm vis}}
   P_l\left( \cos \theta_{ij} \right) \; , \; \; \; \; \;
   l = 1, \, 2, \, 3, \, ...
\end{equation}
the Fox-Wolfram moment ratios can be expressed as $H_l/H_0$, where $\theta_{ij}$ is the opening angle between particles $i$ and $j$, $E_{\rm vis}$ is the total visible energy of the event, and $P_l(x)$ are the Legendre polynomials. 

Figures~\ref{fig:eventshape} and~\ref{fig:eventfox} show the distribution of the event shape variables for 5 and 8 TeV black holes with $n=2 - 6$. The distributions are relatively similar for higher values of $n$, making it hard to distinguish them from one another. Events are relatively spherical with a high transverse energy of a few \Mpl{}.  For higher dimensions, the events become significantly less spherical and are more susceptible to variations in the treatment of the remnant decay.

\FIGURE{
\subfig{a}{\epsfig{file=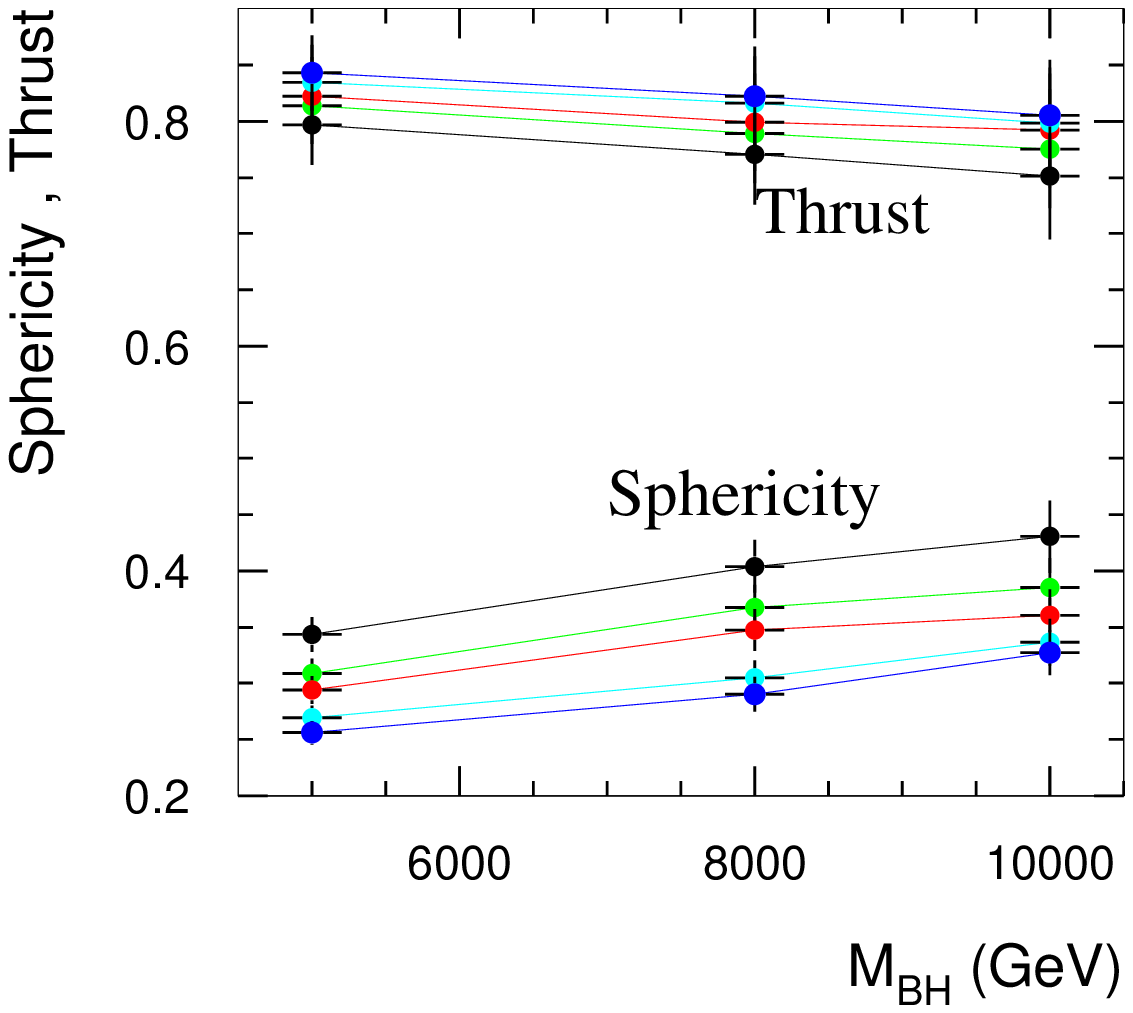, width=7cm}}
\subfig{b}{\epsfig{file=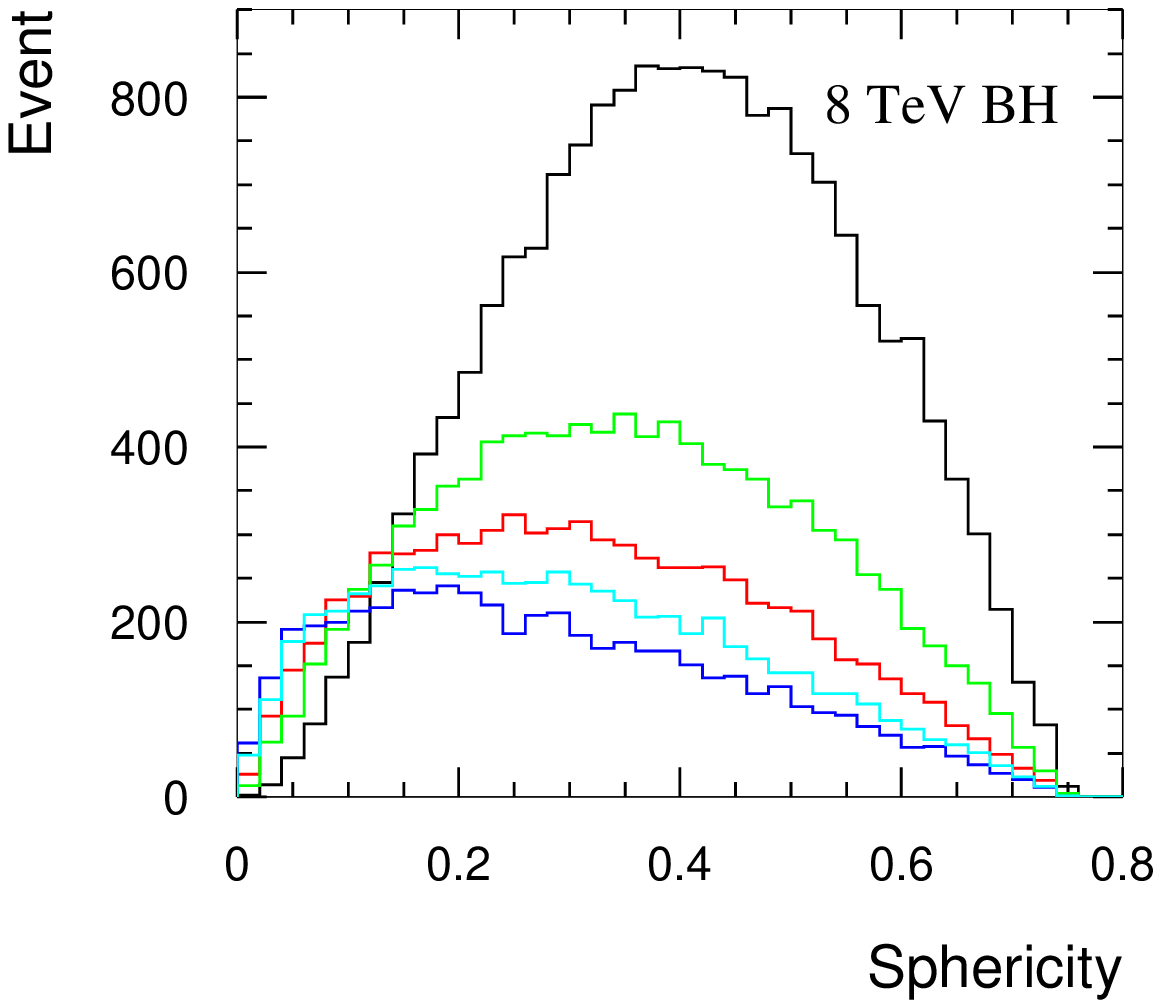, width=7cm}}
\caption{Event thrust and sphericity for $n=2, \, 3, \, 4, \, 5,$ and 6 (black, green, red, cyan, and blue curves, respectively). (a) The variation of the average thrust and sphericity with mass and $n$.  (b) A typical set of distributions of the sphericity for 5~TeV black holes.
}
\label{fig:eventshape}
}

\FIGURE{
\subfig{a}{\epsfig{file=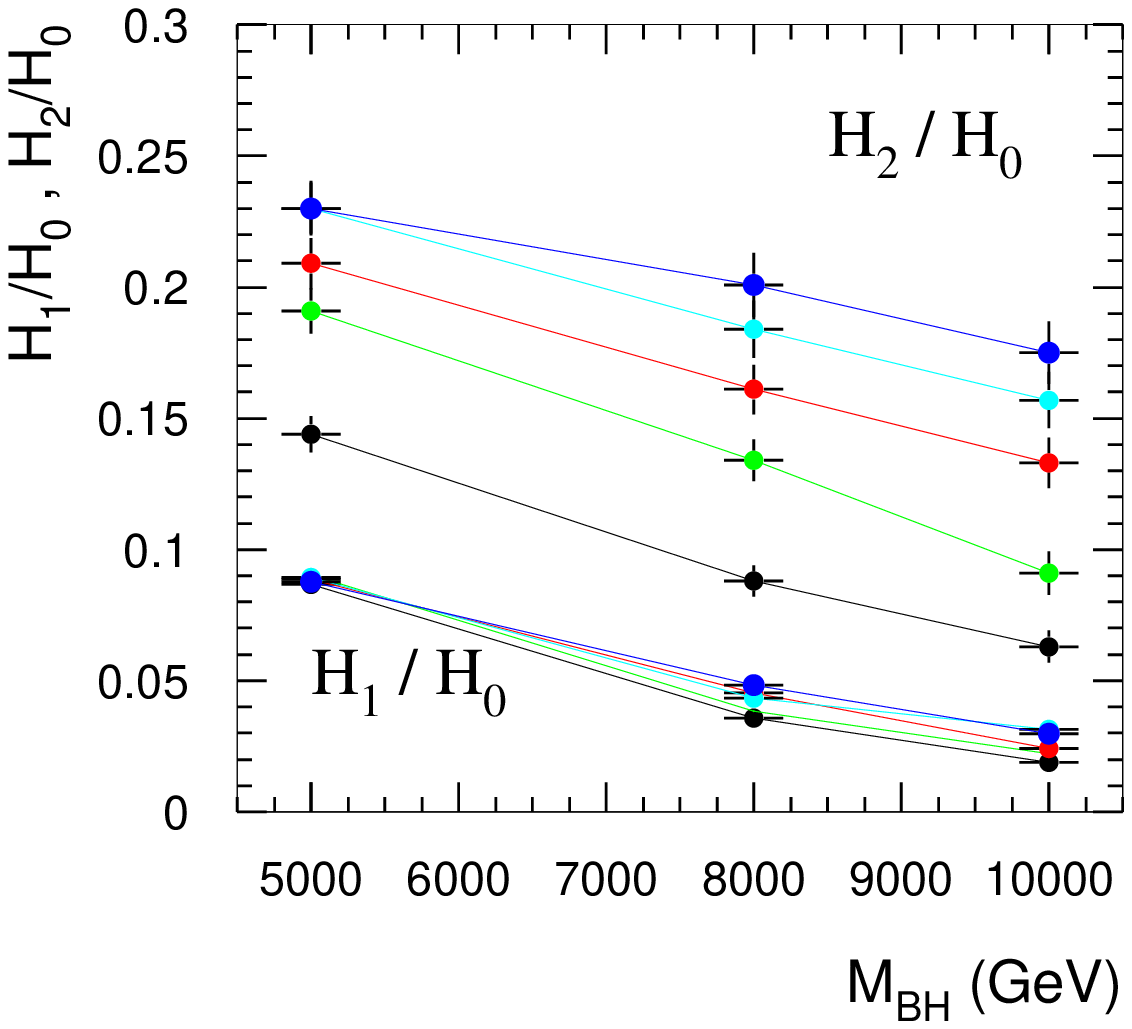, width=7cm}}
\subfig{b}{\epsfig{file=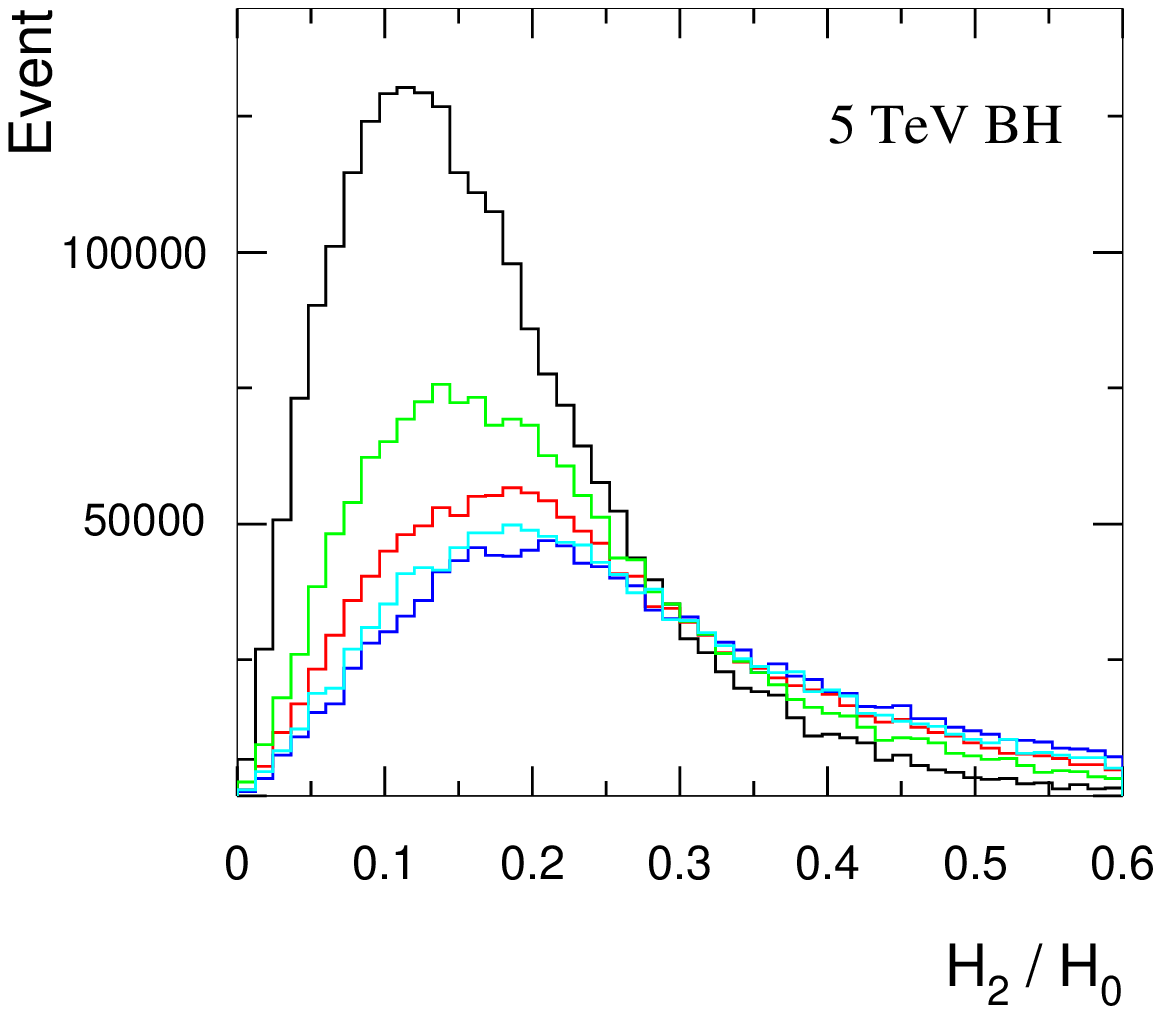, width=7cm}}
\caption{Fox-Wolfram variables for $n=2, \, 3, \, 4, \, 5,$ and 6 (black, green, red, cyan, and blue curves, respectively). (a) The variation with mass and $n$.  (b) A typical set of distributions for 5~TeV black holes.
}
\label{fig:eventfox}
}

\section{Measurement of the black hole mass}
\label{sec:mass}

The black hole 4-momentum is reconstructed simply by summing the 4-momenta of all of the particles in the event.  We have illustrated this procedure for selected black hole mass points 5 and 8~TeV with $n = 2$ to 6 which were generated with masses ranging 200~GeV above and below the selected mass point.  Events were selected by requiring at least 4 jets, all within the acceptance of the ATLAS tracking detector ($|\eta|<2.5$). Requiring the multiplicity to be greater than 4 ensures that the remnant decay will not be too dominant. The transverse momenta of the three highest $p_T$ jets was required to be above 500, 400, and 300~GeV respectively.\footnote{A reconstructed jet was required to have a minimum momentum of 10~GeV within an $\eta-\phi$ cone of radius~0.4.} In order to improve the reconstructed mass resolution, events were rejected if the missing transverse momentum was greater than 100~GeV.
 
The reconstructed Gaussian mass resolution and the overall signal efficiency (the fraction of accepted events) after the selection cuts for 5 and 8~TeV black hole in $n = 2$, 4 and 6 are given in table~\ref{tab:efftable} with sample plots in figure~\ref{fig:massRes}. The mass resolution can be improved slightly by raising the threshold of the jet $p_T$, but at the cost of a sharp drop in overall signal efficiency.

\TABLE{
\renewcommand{\arraystretch}{1.2}
\begin{tabular}{|l|c|c|c|}
  \hline
  \multicolumn{2}{|c||}{Topology} & Mass Resolution (GeV) & Efficiency (\%)
  \\ \hline \hline
                     & $n = 2$ &  202.1  & 26.1  \\ \cline{2-4}
   5~TeV black hole  & $n = 4$ &  188.4  & 30.0  \\ \cline{2-4}
                     & $n = 6$ &  184.4  & 31.9  \\ \hline
                     & $n = 2$ &  293.9  & 13.2  \\ \cline{2-4}
  8~TeV black hole   & $n = 4$ &  234.0  & 17.8  \\ \cline{2-4}
                     & $n = 6$ &  226.4  & 19.3  \\ \hline
\end{tabular}
\caption{\label{tab:efftable}The reconstructed Gaussian mass resolution and the overall signal efficiency after the selection cuts.}
}

\FIGURE{
\subfig{a}{\epsfig{file=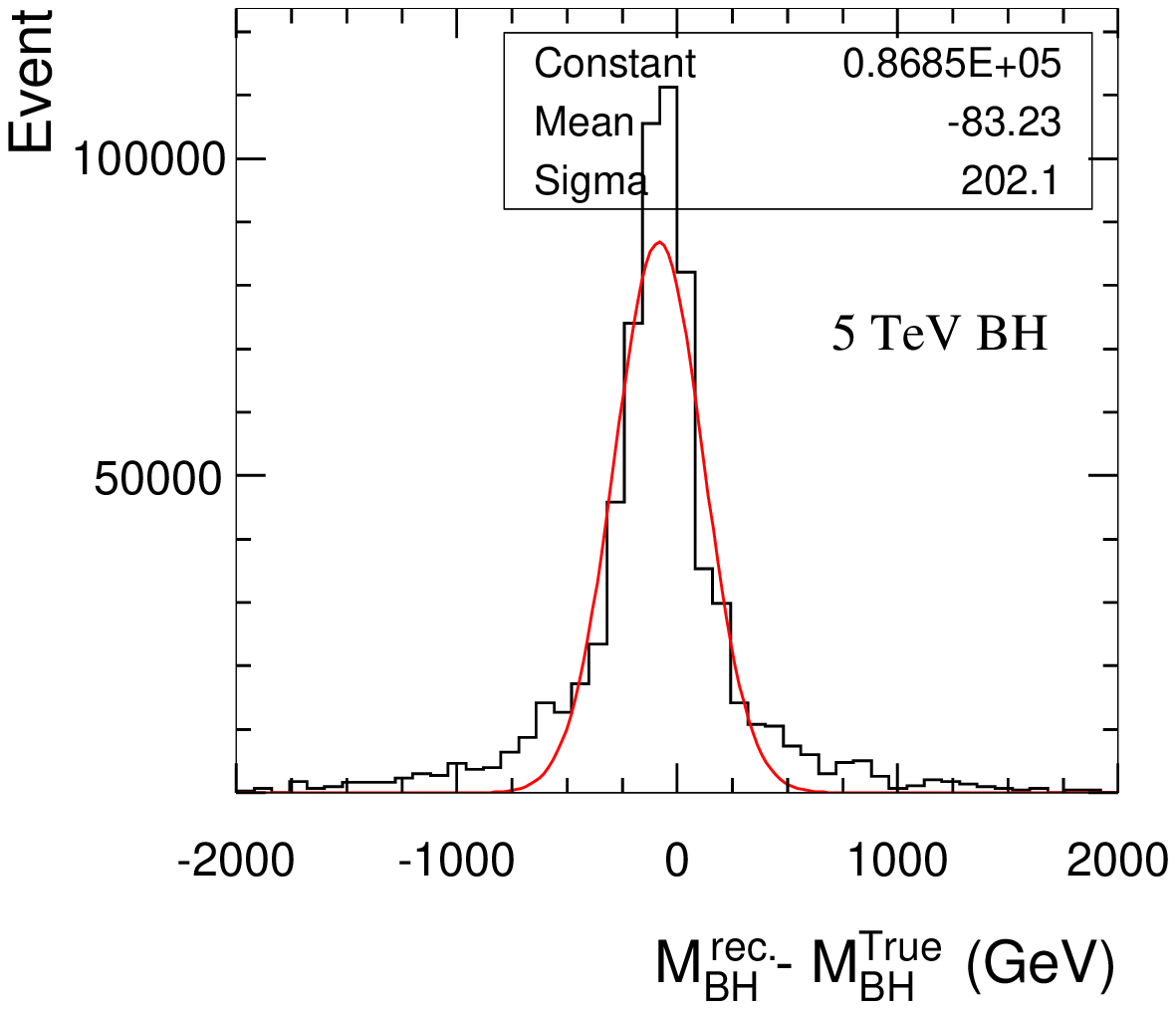, width=7cm}}
\subfig{b}{\epsfig{file=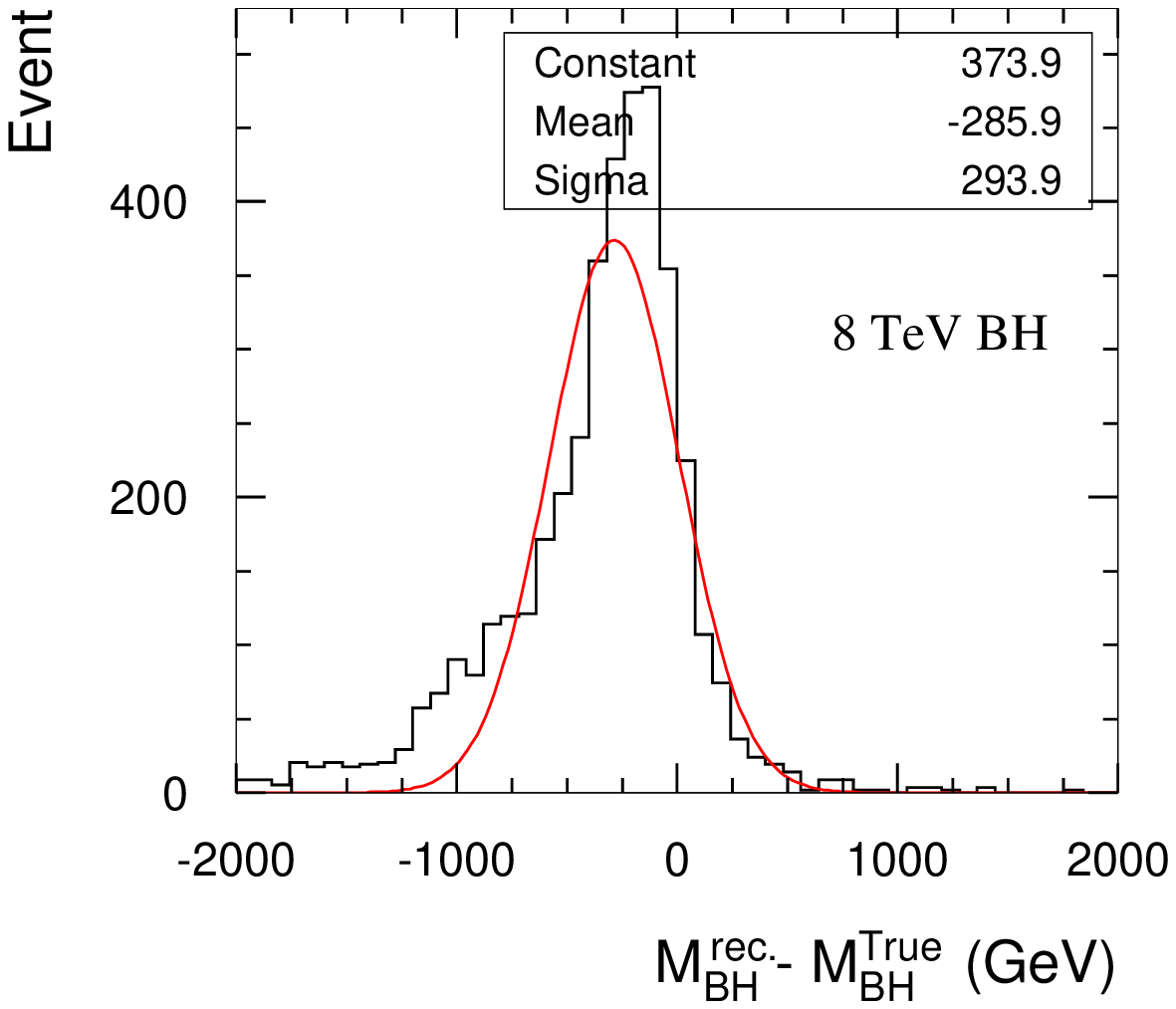, width=7cm}}
\caption{Mass resolution for $n=2$ and (a) $\Mbh=5$~TeV and (b) $\Mbh=8$~TeV.}
\label{fig:massRes}
}

\section{Measurement of the Planck mass}
\label{sec:MPlanck}
Some authors~\cite{Dimopoulos:2001hw} have suggested that since $n$ can be determined from the $T_H$--\Mbh{} relationship (equation~\ref{eq:TempMass}), \Mpl{} can be measured from the normalisation of the temperature.  For reasons outlined in the next section, we choose not to use this method but instead to follow the suggestion of~\cite{ATL-PHYS-2003-037} and determine \Mpl{} from the cross section.  In the convention used in this paper (and also in \cite{ATL-PHYS-2003-037}), the cross section is largely independent of $n$.  Figure~\ref{fig:CrossSection} shows the parton-level cross section including the corrective form factors calculated in~\cite{Yoshino:2002tx}.  As can be seen, there is very little variation with $n$.

\FIGURE{
\epsfig{file=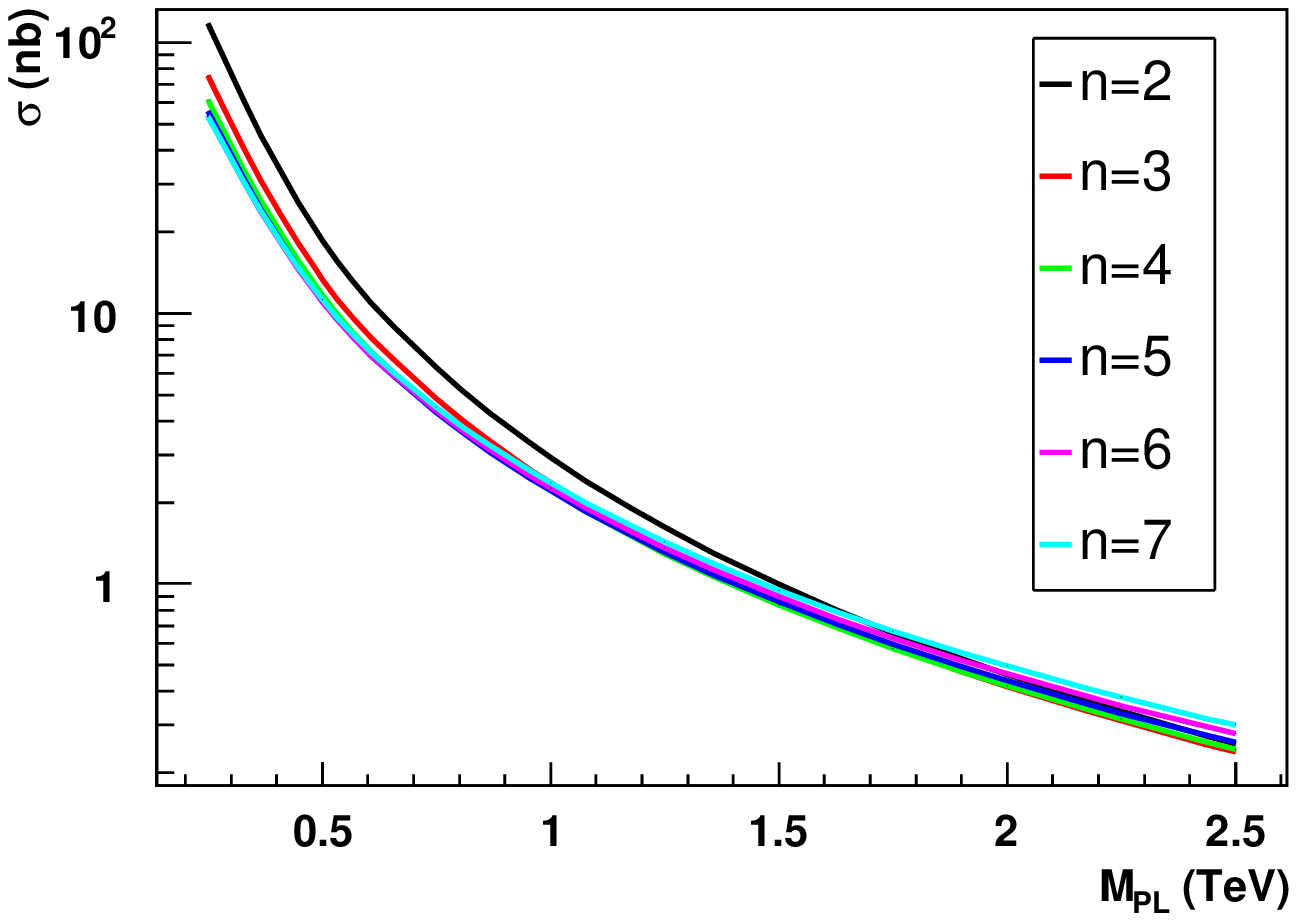, width=10cm}
\caption{The parton level cross section for 5~TeV black holes as a function of \Mpl{} for different values of $n$.  The form factors calculated in~\cite{Yoshino:2002tx} are included.}
\label{fig:CrossSection}
}

Due to the very high statistics available, the measurement of the parton-level cross section will be dominated by the various systematic errors.  The main experimental error will be the luminosity which should be measured to 5\% or better, together with some uncertainty in the efficiency.  This is however likely to be small compared to the theoretical uncertainties discussed in section~\ref{sec:CrossSection}.  We therefore conservatively estimate that the parton-level cross section could be determined to 20\% which, for our test case of $\Mpl=1$~TeV, gives an error in \Mpl{} of about 10\%.  Obviously, the optimal approach is to fit the cross section and temperature data simultaneously and this will be demonstrated at the end of section~\ref{sec:KLAnalysis}.  It is also possible that other processes and observations of new physics at the Planck scale may provide independent measurements of \Mpl{}.

\section{Determination of the number of extra dimensions}
\label{sec:ExtraD}
Measuring the number of extra dimensions is not a straight-forward task given the uncertainties outlined in section~\ref{sec:ModelUncertainties}.  One technique that has been suggested~\cite{Dimopoulos:2001hw} uses the energy spectrum of electrons and photons below $\Mbh/2$.  However the authors of~\cite{Dimopoulos:2001hw} ignore the likely effects on the low energy spectrum from the initial parts of the decay (section~\ref{sec:EarlyStages}), the effect of the remnant decay (section~\ref{sec:RemnantDecay}) and the recoil of the black hole (section~\ref{sec:TimeVar}).  Their analysis is particularly sensitive to these effects because they were attempting to use the variation of $T_H$ with \Mbh{} to measure $n$.  To give some numerical estimates, the expected variation, $T_H(10~\text{TeV})-T_H(5~\text{TeV})$ is about 40~GeV for $n=2$ and 20~GeV for $n=5$ given $T_H(5~\text{TeV}) \sim 200$ GeV.  This is clearly not a large variation to attempt to measure.

It should also be noted that the differences between different number of dimensions become less significant for higher $n$ due to the power law nature of equation~\ref{eq:TempMass}.  We have investigated a number of variables and techniques whilst considering the effects of the many uncertainties.

\subsection{Correlations}
Notwithstanding the comment above, we initially tried to make maximum use of the $\Mbh$--$T_H$ relationship.  In the case that $T_H$ varies with time, there is one $\Mbh$--$T_H$ point per emission, rather than just one per event.  We therefore developed a technique that used all the information.  The method is:
\begin{enumerate}
\item{Reconstruct the black hole from all the particles in the event.}
\item{Determine the first(next) particle to be emitted.}
\item{Use the measured properties of this particle to determine the temperature of the black hole.}
\item{Record this mass--temperature point.}
\item{Reconstruct the black hole for the next stage using all the particles except for those that have been emitted}
\item{Repeat steps 2--5 until there are no particles left.}
\end{enumerate}
There are two key parts to this algorithm: determining the order of the emitted particles and using the particle properties to determine the temperature.  It was hoped that a method might be found which when averaged over many events would give the correct $\Mbh$--$T_H$ relation.

The method used to determine the order was to assume that the softest particles were emitted first.  This is because as the black hole decays it gets hotter, and so the average energy of the emitted particles should increase.

The most probable energy and thus $p_T$ for the emitted particle is proportional to the temperature of the black hole.  Since the expected $\Mbh$--$T_H$ relation is a power law, the $\Mbh$--$p_T$ relation should have the same dependence on $n$.  Thus we have plotted the average $p_T$ of the emitted particles for each mass bin.

Figure~\ref{fig:corrSim} shows the result of this method.  The shape of the graph is very different from that expected, but does show a separation between the numbers of dimensions.  This is perhaps not surprising since all black hole decays will at some point be affected by the kinematic limit problem to which this technique is particularly sensitive.  Unfortunately this means that extracting the number of dimensions is extremely difficult to do without relying purely on Monte Carlo data.  Further details on this method are given in~\cite{Palmer:Thesis}.

\FIGURE{
\subfig{a}{\epsfig{file=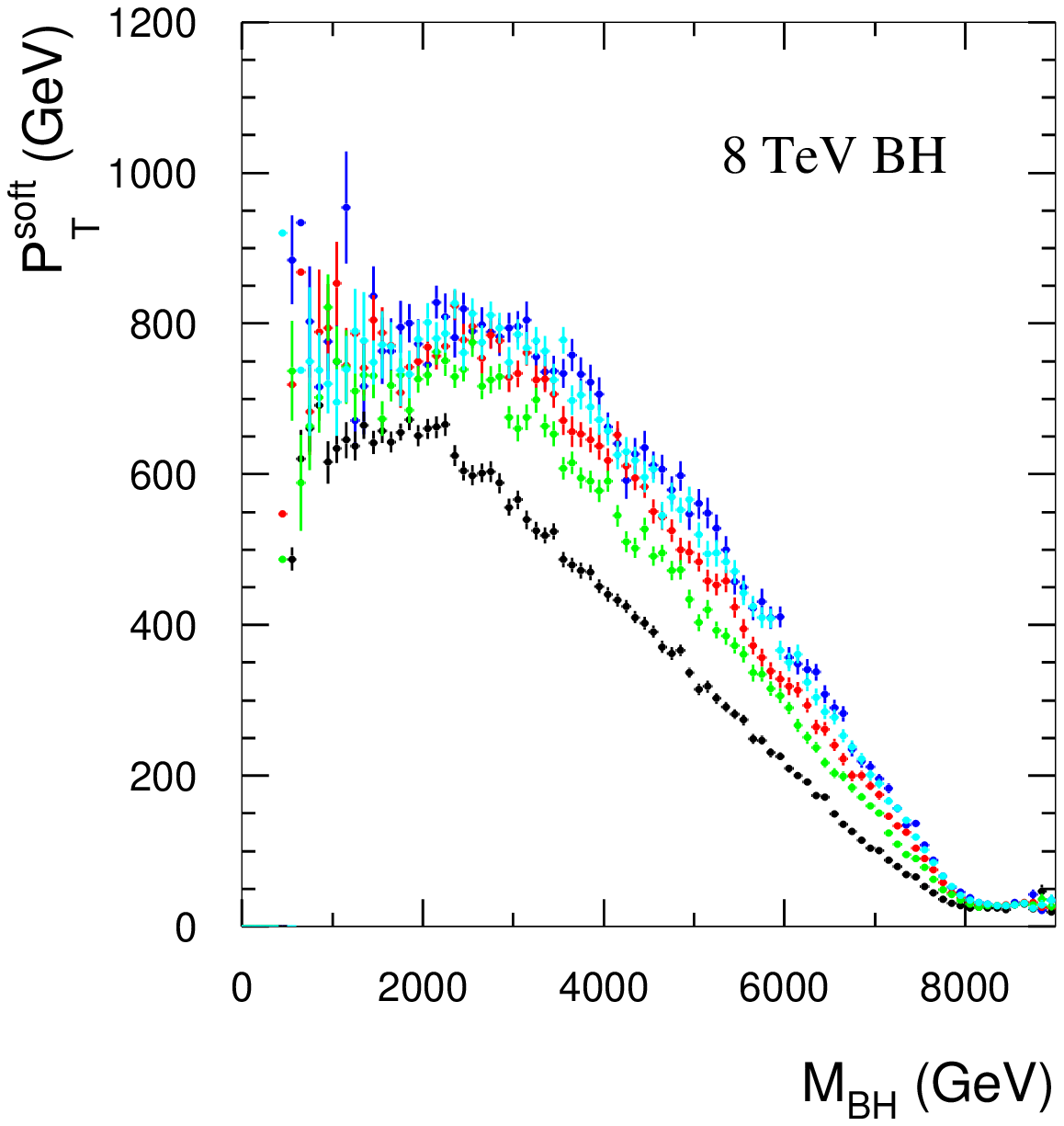, width=7cm}}
\subfig{b}{\epsfig{file=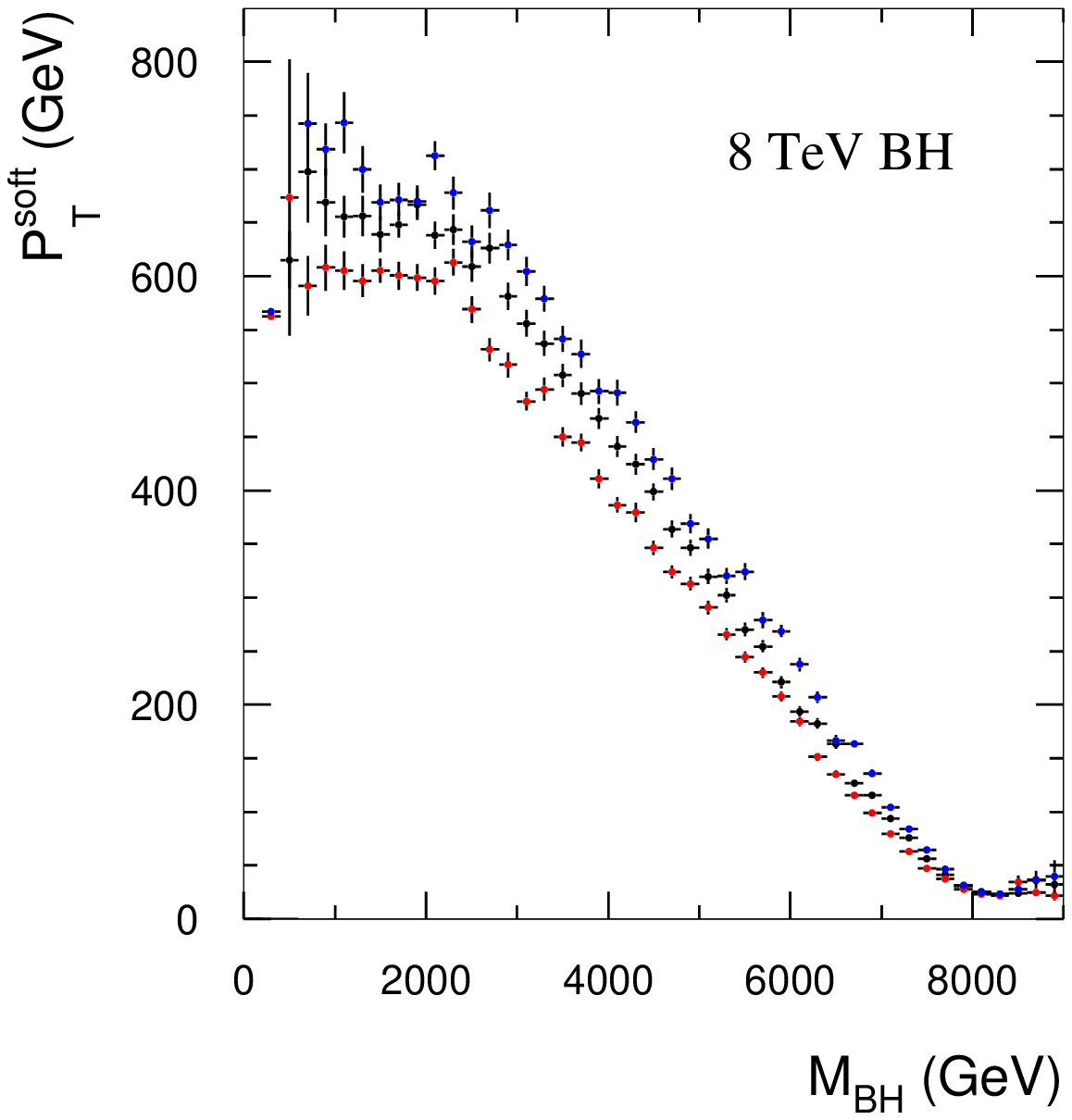, width=7cm}}
\caption{Correlation plots (see text) for 8~TeV black holes (a) with $n=$2, 3, 4, 5, 6 and (b) with \Mpl{}=900, 1000, 1100~GeV.}
\label{fig:corrSim}
}

\subsection{Kinematic limit}
\label{sec:KLAnalysis}

In this section we present a new idea that shows a strong variation with the number of dimensions and is valid for many different scenarios. It may therefore allow $n$ to be determined despite the many uncertainties that affect black hole decays.

If a particle is emitted with an energy close to the kinematic limit (i.e. $E\sim \Mbh/2$), then that particle is probably the first to be emitted.  In particular, it is possible to measure the fraction of events, $p$, where the highest energy particle has an energy, $E_{max} > E_{cut}$ where $E_{cut} = \Mbh/2 - E_d$ and $E_d$ is a parameter of the analysis that can be chosen, but should be small.  The probability of the first emission being greater than $E_{cut}$ can also be calculated from integrating the Planck spectrum (equation~\ref{eq:Spectrum}) thus there is a reasonably direct connection between the experimental measurement and theory.

This method has many advantages: since it deals with the first emission, the probability of $E_{max}>E_{cut}$ will be the same regardless of whether the black hole temperature is time-varying or not.  Also, since we restrict ourselves to black hole with masses much larger than the Planck mass, this measurement will not be affected by Planck scale effects or the remnant decay.  Any effects which modify the low energy part of the spectrum should also have no effect.  Finally, the effect of the black hole boost can be taken into account by determining $E_{max}$ in the black hole rest frame.

This technique is however strongly affected by the uncertainty in dealing with the kinematic limit (see
section~\ref{sec:KinematicLimit}) since we do not know what the shape of the energy spectrum would be near the kinematic limit.  Fortunately this uncertainty can be controlled by putting upper and a lower bounds on the theoretical estimate of $p$.  The lower bound is set by assuming that none of the unphysical region of the Planck spectrum corresponds to an emission with energy greater than $E_{cut}$ in the actual energy spectrum.  The lower bound is therefore
\begin{equation}
p_{lower} = k\int^{\Mbh/2}_{E_{cut}} P(E) dE
\end{equation}
where $k$ is a normalising constant, $P(E)$ is the Planck spectrum (equation~\ref{eq:Spectrum}) and we have ignored the effects of the `grey-body' factors (which are largest at low energy).  The difference between fermions and bosons is small here. The upper bound uses the opposite assumption to the lower bound: that is, that all of the unphysical region in the Planck spectrum would correspond to an emission with energy greater than $E_{cut}$ in the actual energy spectrum.  The upper bound is therefore
\begin{equation}
p_{upper} = k\int^{\infty}_{E_{cut}} P(E) dE\,.
\end{equation}
It should be noted that at very high energies and in the semi-classical limit, $p\rightarrow0$ as $\Mbh\rightarrow\infty$ for any fixed $E_d$.  These equations are therefore only valid when $p$ is small.

There are two competing effects: we would like to set $E_{cut}$ as low as possible so that the upper and lower bound are similar (note that they move apart as $E_{cut} \rightarrow \Mbh/2$).  However, the lower we set the cut, the greater the chance that the highest emission is not the first one.  Whilst this is not in itself a problem, it does mean that the differences between the time varying and non-time varying cases become more pronounced.  In the rest of this study we have set $E_d = 400$~GeV --- setting it lower is difficult as experimental resolution effects start to dominate.  The chance of a soft emission is strongly temperature dependent.  Points have been omitted if this chance is greater than 50\% which affects only $n=2$, 5~TeV black holes, but would be more significant at lower \Mpl{}.

We have included an approximate corrective factor for the emission of a first soft particle.  This was calculated as
\begin{equation}
k_{cor} = k\int^{x E_{cut}}_0 P(E) dE
\end{equation}
giving $p_{cor} = (1+k_{cor}) p$.  It is possible for a first emission of any energy to be followed by an emission of energy up to the kinematic limit, but the probability of this drops sharply if the first emission has an energy greater than $2 E_{cut}$.  The correction factor, $x$, was therefore set to 2 for the lower bound and 3 for the upper bound.

A plot of the upper and lower bounds on $p$ for $\Mpl{}=1$ TeV is shown in figure~\ref{fig:KLBands}.  As can be seen, if \Mpl{} is known, there is good potential for extracting $n$ if it is below 5 and could be constrained if $n$ is larger.

\FIGURE{
\epsfig{file=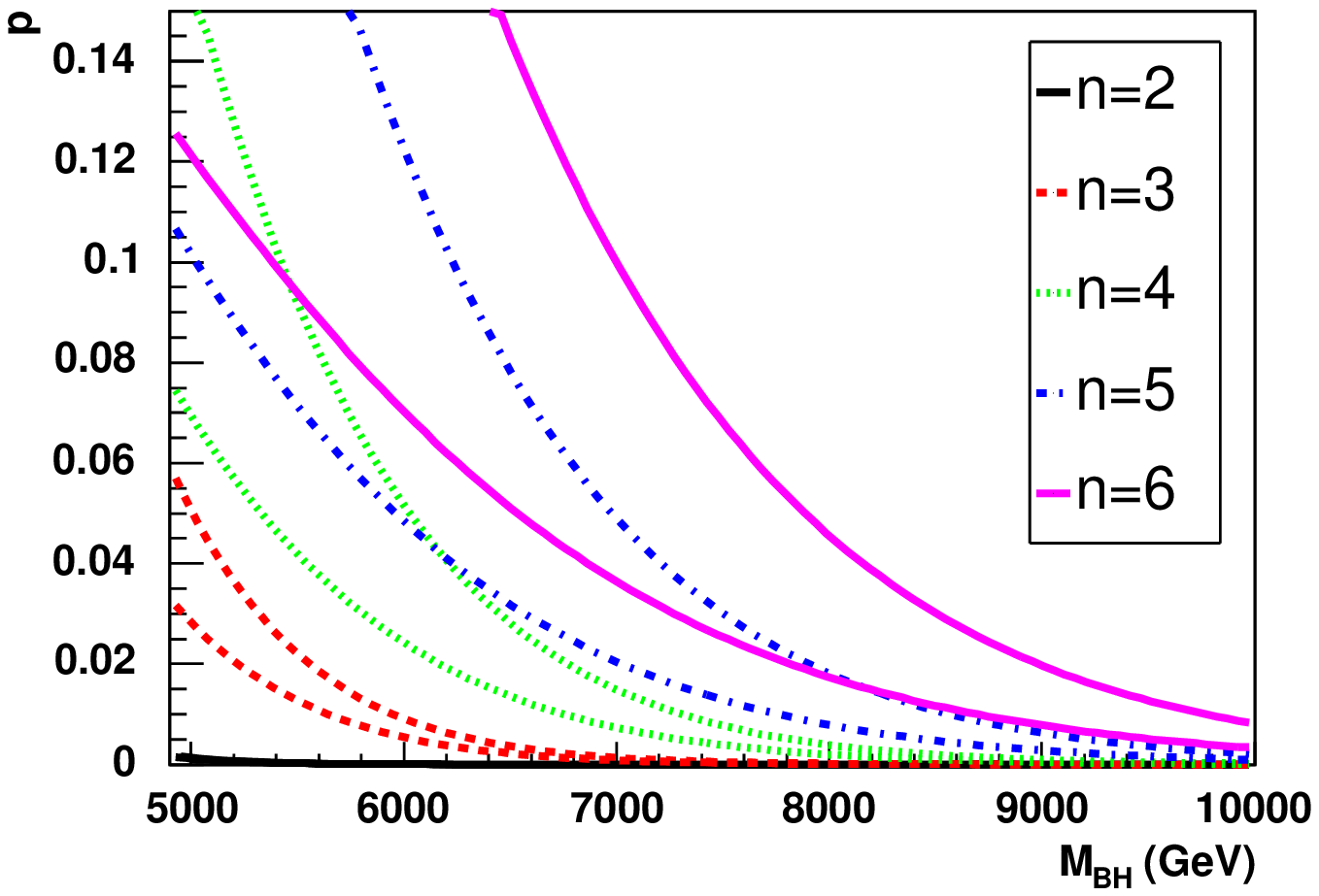, width=0.601\textwidth}
\caption{Upper and lower bounds for $p$ for different $n$ with $\Mpl=1$ TeV.}
\label{fig:KLBands}
}

One experimentally tricky aspect of this measurement is that since the black hole mass is measured by adding all the particles in the event, the maximum energy in the reconstructed black hole rest frame must be less than $\Mbh/2$.  This introduces a bias to low energies in the measurement of $E_{max}$ and thus reduces the measurement of $p$.  This has been corrected for by increasing $E_d$ by 100~GeV which is an estimate for the average mis-measurement of $E_{max}$.  It should be noted that not boosting into the black hole rest frame would give a very significant over-measurement of $p$.  In addition, to ensure that the black hole mass was well measured, events were excluded if they had $\ptmiss{} > 100$~GeV, had any particles with $|\eta|>2.5$ or if the three highest energy particles accounted for more than 95\% of the total energy in the black hole rest frame.

Figure~\ref{fig:KLModelInvariant} shows the results of this analysis for four different models all with $n=4$.  The theoretical upper and lower limits are also shown.  For models which have the kinematic cut off (see section~\ref{sec:KinematicLimit}), we would expect them to be consistent with the lower limit; this is the case for plots a, b and d in figure~\ref{fig:KLModelInvariant}.  However, if the kinematic limit is on and the remnant decay is set to 2-body (plot c), we would expect the data to be consistent with the upper limit.  This is because if an energy is chosen in the forbidden region (which is the additional region included in the upper limit integral), it will cause the decay to be terminated and the black hole to split into two.  One of the two remnant decay particles must pass the cut.  As can be seen from the figure, the plots do agree with these expectations.

\FIGURE{
\subfig[ Test case (see section~\ref{sec:event})]{a}{\epsfig{file=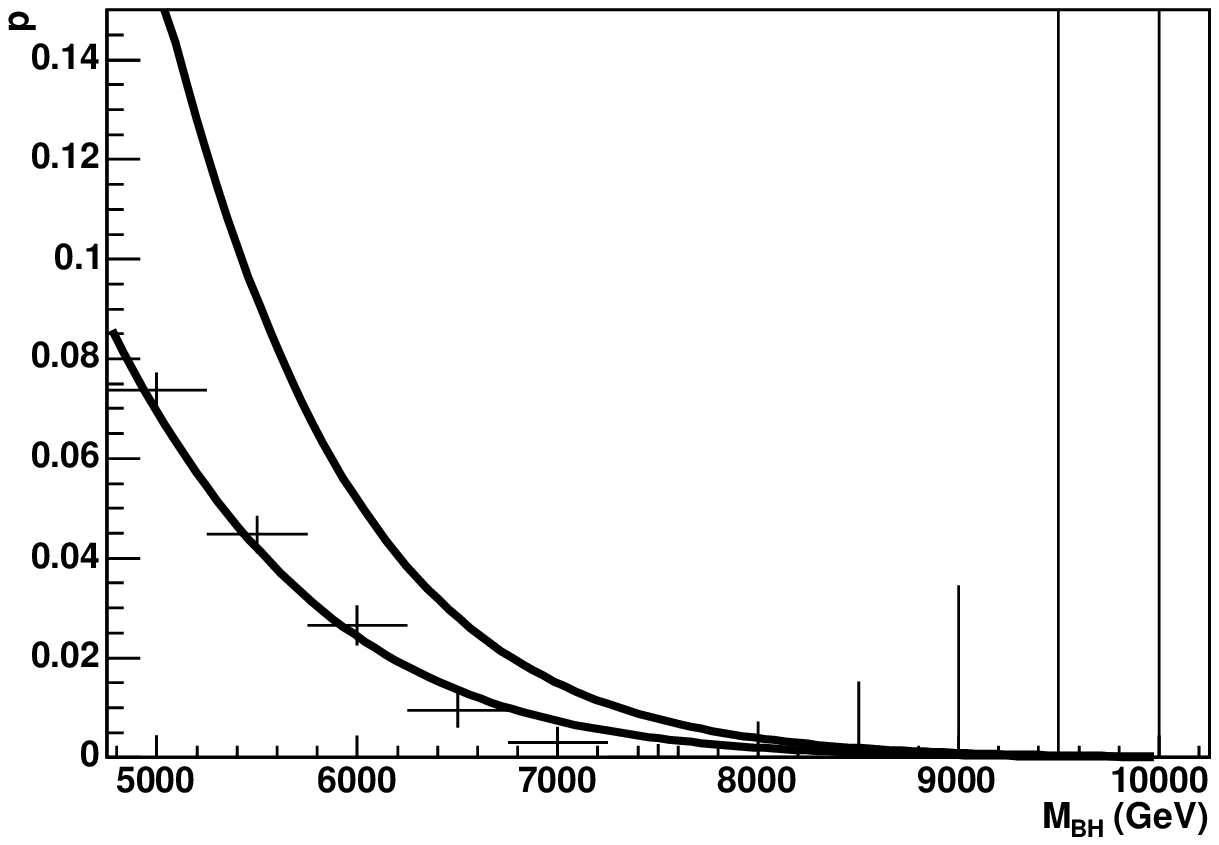, width=7cm}}
\subfig[ No time variation]{b}{\epsfig{file=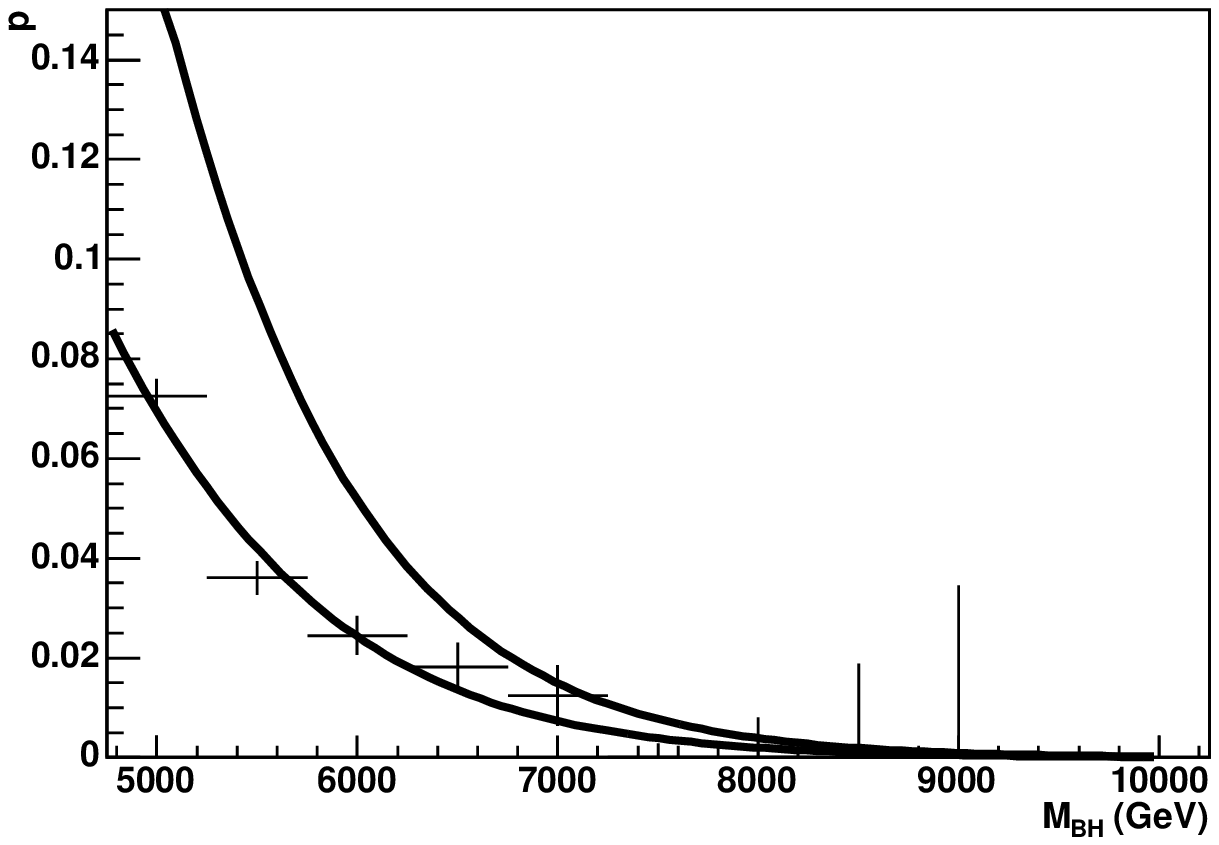, width=7cm}}
\subfig[ Kinematic cut on]{c}{\epsfig{file=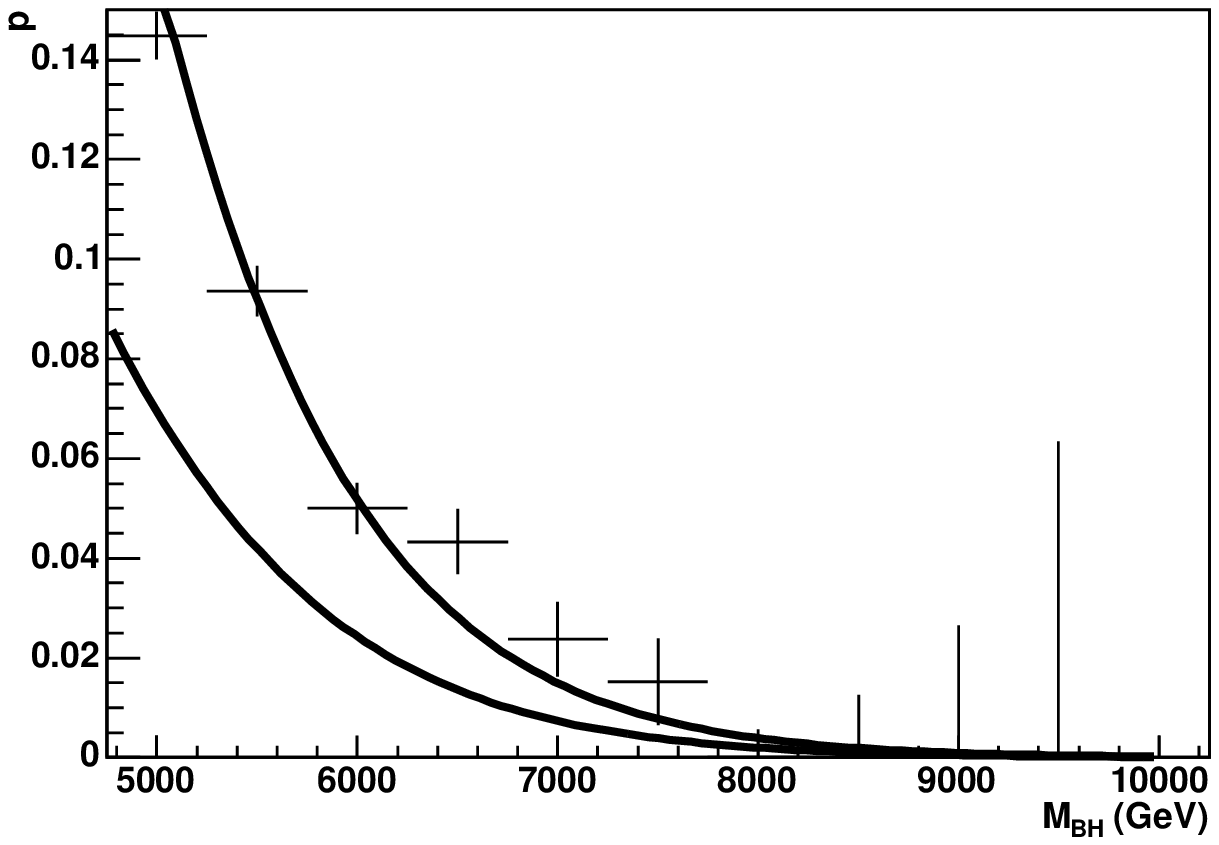, width=7cm}}
\subfig[ 4-body remnant decay]{d}{\epsfig{file=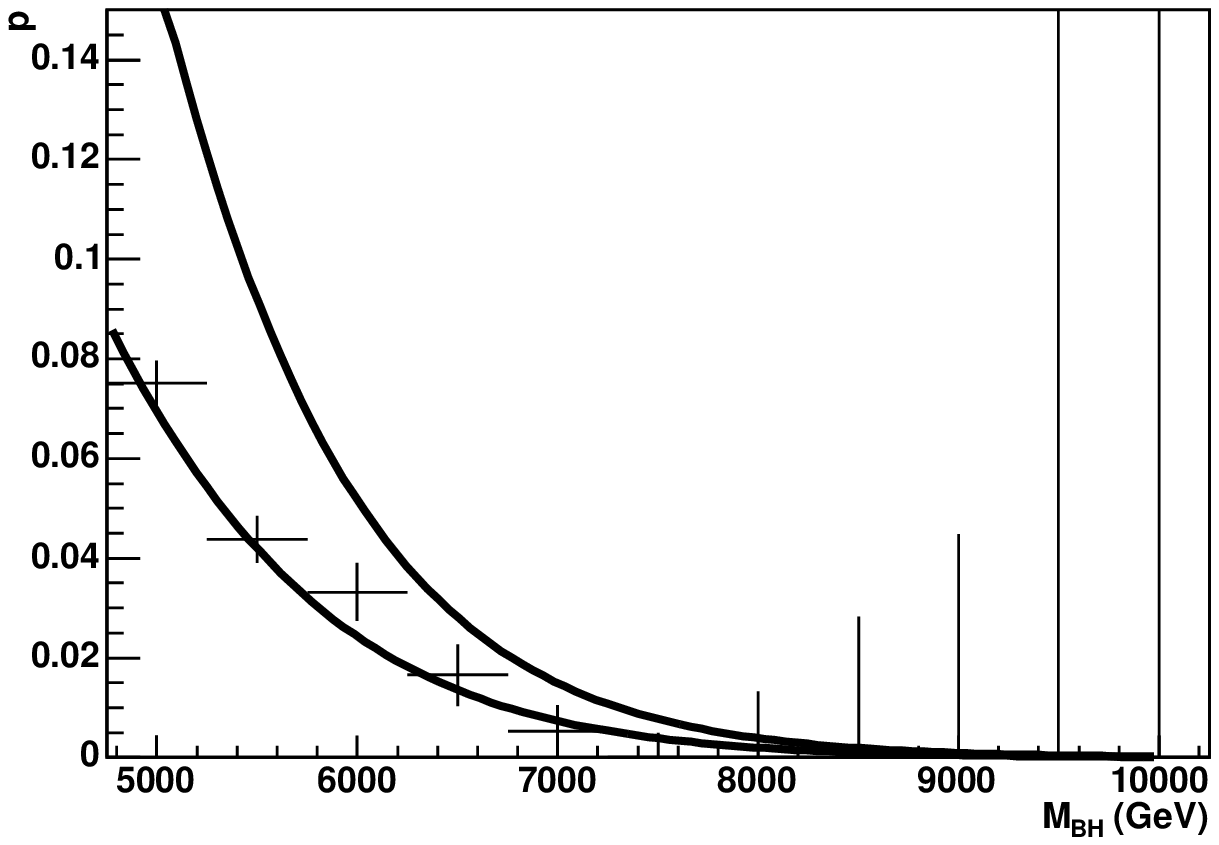, width=7cm}}
\caption{Fraction of events passing the cut, $p$, as a function of \Mbh{} for different models all with $n=4$.  Upper and lower bounds for $n=4$ are also shown.}
\label{fig:KLModelInvariant}
}

Figure~\ref{fig:KLNDependant} shows similar plots to figure~\ref{fig:KLModelInvariant}, but for the test case (see section~\ref{sec:event}) with different values of $n$.  This emphasises that this technique is sensitive to $n$ whilst being largely model independent.  Note that at high $n$, the data start to drop below the lower limit.  This is due to the high temperature here ($T_H\sim470$ GeV for $n=5$) which significantly reduces the multiplicity.  This result suggests that this analysis has an upper limit of validity in the region of $T_H\sim450$--500~GeV.  Also, from figure~\ref{fig:KLNDependant}a, it can be seen that at low temperatures, this analysis will not measure the temperature (unless lighter black holes are seen, or very high statistics are available), but would instead place an upper limit on it.  This may be enough to constrain $n$, or alternatively, measuring the energy distribution may be more successful here and these analyses could be combined to give a temperature measurement.

\FIGURE{
\centering
\subfig[ $n=2$]{a}{\epsfig{file=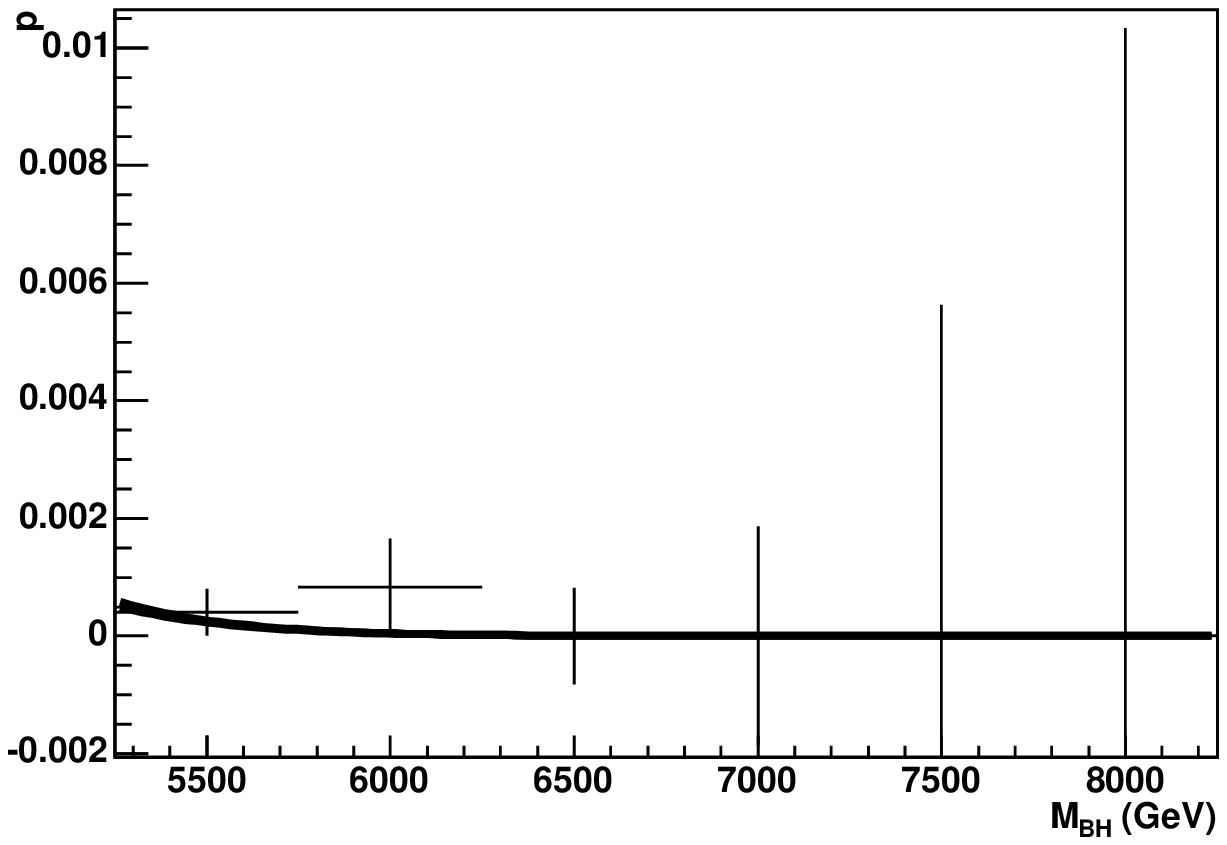, width=7cm}}
\subfig[ $n=3$]{b}{\epsfig{file=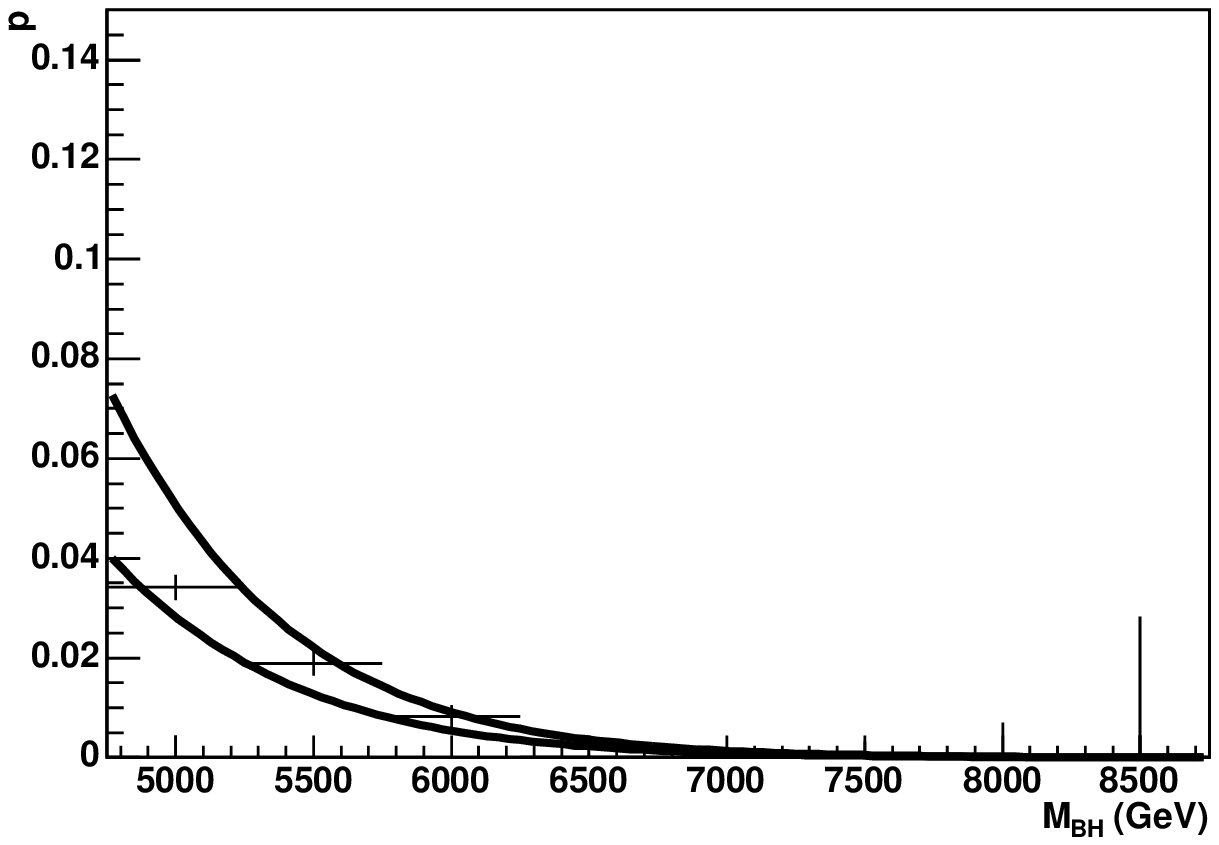, width=7cm}}
\subfig[ $n=4$]{c}{\epsfig{file=KL_n4_Std.eps, width=7cm}}
\subfig[ $n=5$]{d}{\epsfig{file=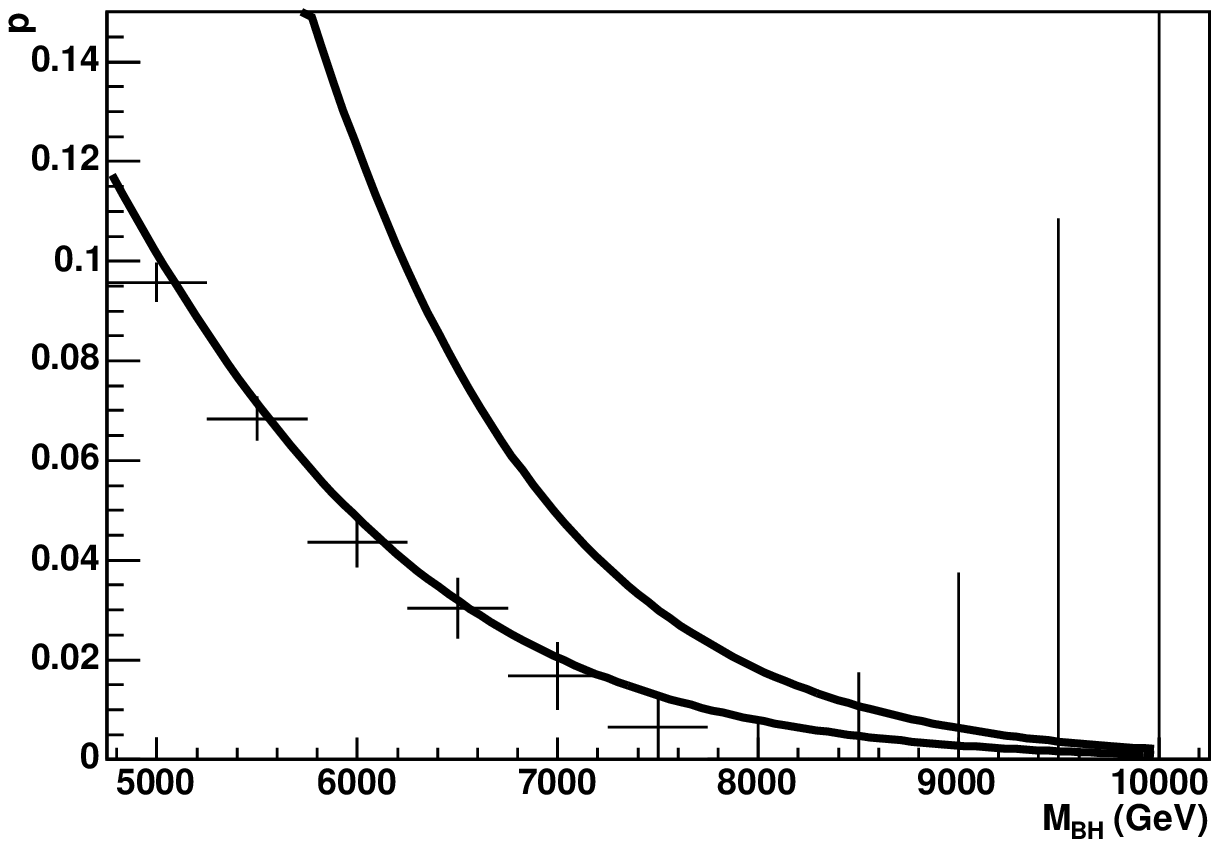, width=7cm}}
\caption{Fraction of events passing the cut, $p$, as a function of \Mbh{} for different values of $n$ for the test case.  Appropriate upper and lower bounds are shown.}
\label{fig:KLNDependant}
}

At higher temperatures, this technique becomes strongly sensitive to the temperature of the black hole.  Indeed the plots above can easily be converted into temperature as a function of \Mbh.  This has been done for the test case with $n=4$ in figure~\ref{fig:KinematicLimitTemp} which also includes a band equivalent to the upper and lower limits on $p$.  The band includes a systematic uncertainty on the measurement of the black hole mass of $\pm200$~GeV.  Note that as suggested at the beginning of this section, the temperature variation with \Mbh{} is not determined sufficiently to constrain $n$ without first measuring \Mpl{}.  So instead, we fix the normalisation of $T_H$ at the black hole mass at which it is best measured.  In this case, we take $T_H=340\pm30$ GeV at $\Mbh=7000$ GeV.  This measurement has been taken together with the parton-level cross section with an error of 20\% (see section~\ref{sec:MPlanck}) and used to determine the model parameters $n$ and \Mpl{} in figure~\ref{fig:KinematicLimitResult}.  In this case, this gives an error on the determination of $n$ of 0.75 and an error on \Mpl{} of 150~GeV.  These results are indicative of how well this analysis can do.  If the cross section error were reduced to 10\%, the error on \Mpl{} would be 70~GeV and on $n$, 0.6, showing that, as expected, the cross section dominates the determination of \Mpl{}.

\DOUBLEFIGURE{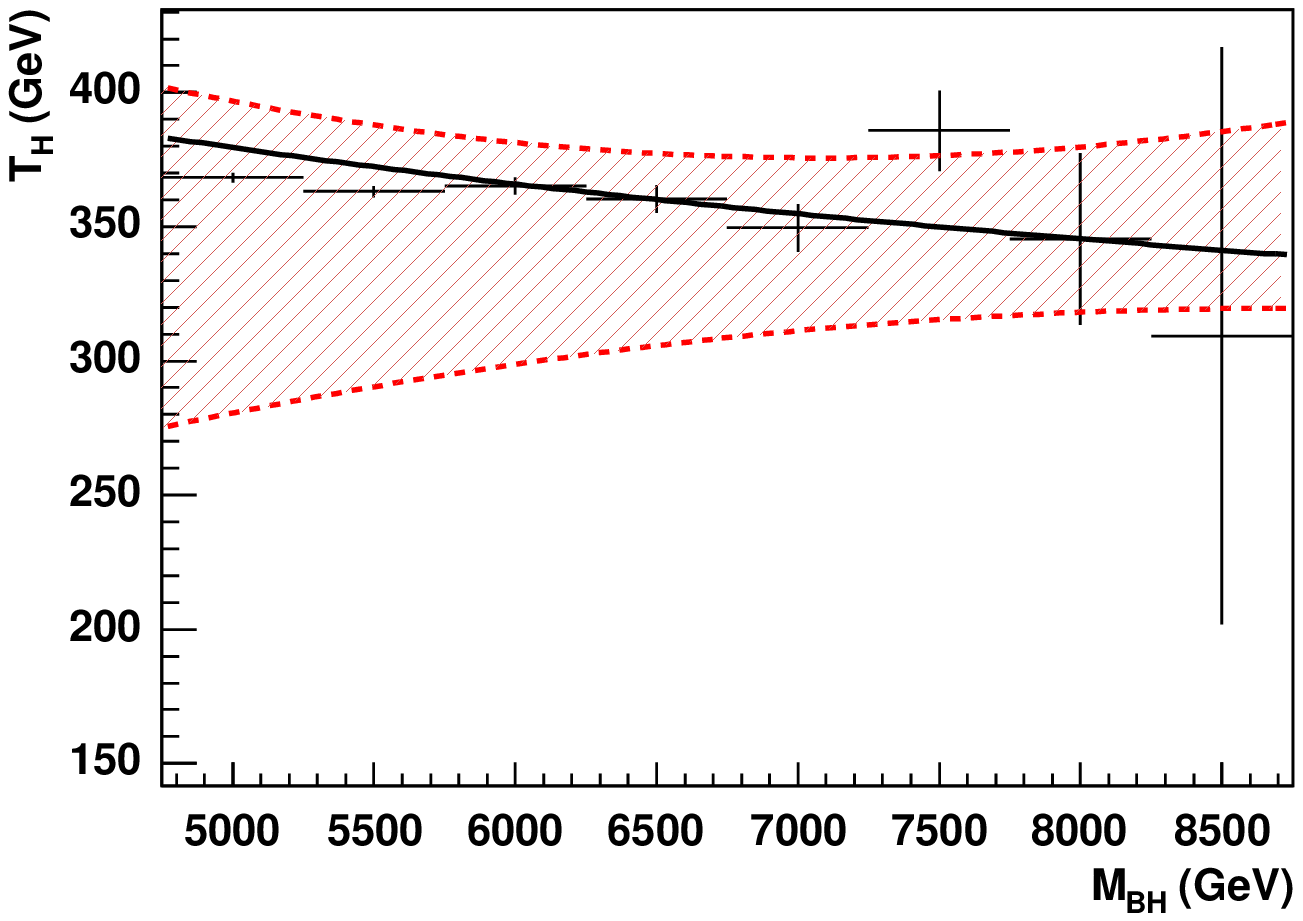, width=7cm}{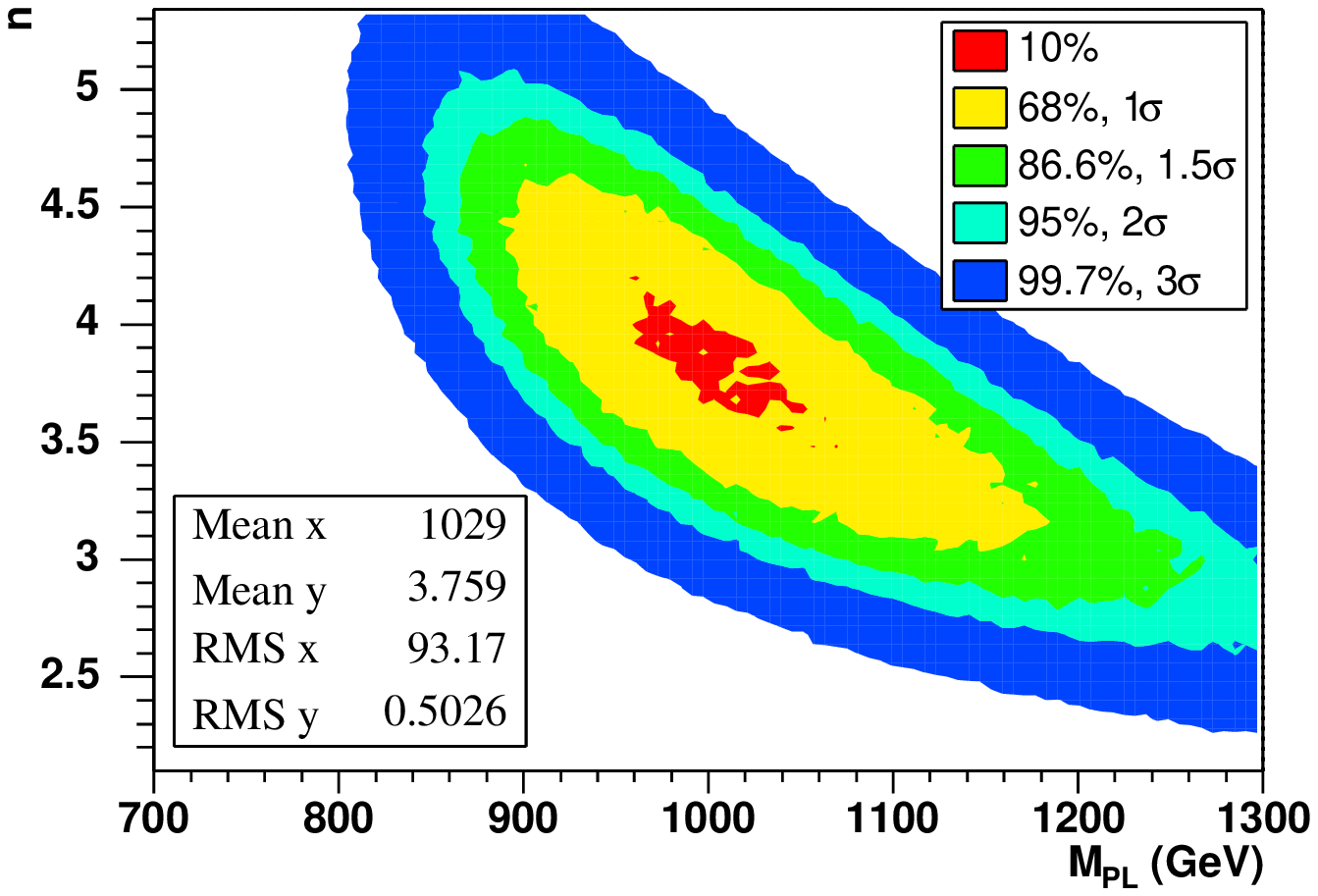, width=7cm}{\label{fig:KinematicLimitTemp}Temperature against \Mbh{} for $n=4$ and 30~fb$^{-1}$ of integrated luminosity.  The band shows the systematic uncertainty corresponding to the upper and lower bounds on $p$ with a systematic on the \Mbh{} measurement of $\pm200$~GeV.}{\label{fig:KinematicLimitResult}The determination of $n$ and \Mpl{} from the measurement of $T_H$ and an assumed measurement of the parton-level cross section (see text).}

Any improvement in our understanding of how the distribution above the kinematic limit should be handled, or how the remnant would decay, would greatly improve this analysis by reducing the width of the bands in figures~\ref{fig:KLModelInvariant},~\ref{fig:KLNDependant} and~\ref{fig:KinematicLimitTemp}.

\section{Conclusions}
We have discussed the many theoretical uncertainties that can affect black hole decays and shown that in at least one case, these can lead to systematic mis-measurements of the number of extra dimensions if the analyses previously suggested are used.  We have then shown the characteristics of black hole decays as they would be measured in the ATLAS detector.  A number of different attempts to determine the model parameters have been discussed and a new technique has been introduced.  This new technique has been shown to control many of the theoretical uncertainties and can be used to measure the black hole temperature.  We have applied this technique to our test case with four extra dimensions and found the temperature to be $340\pm30$ GeV for a black hole mass of 7~TeV.  This was combined with the parton-level cross section, assumed to be known to 20\%, to give estimates of the Planck mass and the number of extra dimensions.  We conclude that in this case the Planck mass can be determined to 15\% and the number of extra dimensions to $\pm$0.75, with strongly correlated errors.

\acknowledgments
We thank members of the Cambridge SUSY Working Group; in particular C. G. Lester, for many useful suggestions and discussions.  BRW thanks the CERN Theory Group for hospitality during part of this work.  We thank the ATLAS Collaboration for the use the ATLAS physics analysis framework and tools which are the result of collaboration-wide efforts.  This work was funded by the U.K. Particle Physics and Astronomy Research Council.

\bibliography{BH}

\end{document}